\documentclass[aps,pra,preprint,amsmath,amssymb,showpacs]{revtex4}
\usepackage{graphicx,epsfig,epsf}
\usepackage{dcolumn}
\usepackage{bm}

\begin{document}
\newcommand{\ri}{{\rm i}}
\newcommand{\re}{{\rm e}}
\newcommand{\bx}{{\bf x}}
\newcommand{\br}{{\bf r}}
\newcommand{\bk}{{\bf k}}
\newcommand{\bE}{{\bf E}}
\newcommand{\bR}{{\bf R}}
\newcommand{\bn}{{\bf n}}
\newcommand{\rSi}{{\rm Si}}
\newcommand{\beps}{\mbox{\boldmath{$\epsilon$}}}
\newcommand{\rg}{{\rm g}}
\newcommand{\tr}{{\rm tr}}
\newcommand{\xmax}{x_{\rm max}}
\newcommand{\ra}{{\rm a}}
\newcommand{\rx}{{\rm x}}
\newcommand{\rs}{{\rm s}}
\newcommand{\rP}{{\rm P}}
\newcommand{\up}{\uparrow}
\newcommand{\down}{\downarrow}
\newcommand{\hc}{H_{\rm cond}}
\newcommand{\kb}{k_{\rm B}}
\newcommand{\cI}{{\cal I}}
\newcommand{\cE}{{\cal E}}
\newcommand{\cC}{{\cal C}}
\newcommand{\Ubs}{U_{\rm BS}}
\newcommand{\qq}{{\bf ???}}
\sloppy

\title{Interference versus success probability in quantum algorithms with imperfections} 
\author{Daniel Braun and Bertrand Georgeot}
\affiliation{Laboratoire de Physique Th\'eorique, 
  Universit\'e de Toulouse, CNRS,  31062 Toulouse, FRANCE} 

\begin{abstract}
We study the influence of errors and decoherence on both the
  performance of Shor's
  factoring algorithm and Grover's search algorithm, and on the amount of
  interference in these algorithms using a recently proposed interference
  measure. We consider systematic unitary errors,  
  random  unitary errors, and decoherence processes. We show that unitary
  errors which destroy 
  the interference destroy the efficiency of the algorithm, too.
However, unitary errors may also create useless additional interference. 
  In such a case  the total amount of interference can increase, while  the 
  efficiency of the quantum computation decreases. 
 For decoherence due to phase flip errors,
  interference is destroyed for small error probabilities, and converted
  into destructive 
  interference for error probabilities approaching one, leading to
  success probabilities which can even drop below the classical
  value. Our results show that in general interference is necessary in
  order for a quantum algorithm to outperform classical computation, but
  large amounts of interference are not sufficient and can even lead to
  destructive interference with worse than classical success rates.
\end{abstract}
\maketitle
\section{Introduction}
Quantum algorithms differ from classical stochastic algorithms by the facts
that they have access to entangled quantum states and that they can make use
of  
interference effects between different computational paths  
\cite{Nielsen00}. These effects can be exploited for
spectacular 
results. Shor's algorithm factors large integers in a time which is
polynomial in the number of digits \cite{Shor94}, and Grover's search algorithm
finds an item in an unstructured database of size $N$ with only $\sim
\sqrt{N}$ queries \cite{Grover97}. Many other quantum algorithms have
building blocks similar to those 
developed in those two seminal papers
(e.g.\cite{Boyer98,Brassard02,Georgeot04}.  
But more than twenty years after the discovery of the first quantum
algorithm \cite{Deutsch85} it is still not clear what exactly
is at the origin of the speed up of quantum algorithms compared to their
classical counterparts. 
Large
amounts of entanglement must necessarily be generated in a quantum algorithm
that offers an exponential speed-up over classical computation
\cite{Jozsa03}, and 
tremendous effort has been spent to develop methods to detect 
and quantify entanglement in a given quantum state (see
\cite{Bruss01,DeSLS05} for recent reviews). However, the
creation of large amounts of entanglement does certainly not suffice 
for getting
an efficient quantum algorithm, and it remains to be elucidated what are
both necessary and sufficient requirements.

While there seems to be general agreement that interference plays an
important role in quantum algorithms \cite{Cleve98,Bennett00,Beaudry05},
surprisingly, it 
has remained almost unexplored in computational complexity theory. 
Recently we introduced a measure of interference in order to quantify the
amount of interference present in a given quantum algorithm (or, more
generally, in any quantum mechanical process in a finite dimensional Hilbert
space) \cite{Braun06}. It turned out that both Grover's and 
Shor's algorithms use an exponential amount of interference when the entire
algorithm is considered. Indeed, many useful quantum algorithms
start off with superposing coherently all computational basis states at
least in one register,
which is a process that makes use of  massive interference (the number of
i--bits, a logarithmic unit of interference,  equals
the number of qubits of the register \cite{Braun06}). Both algorithms differ
substantially, 
however, in 
their exploitation of interference in the subsequent non--generic part: Shor's
algorithm uses 
exponentially large interference also in the remaining part of the algorithm
due to a quantum Fourier transform (QFT), whereas the remainder of Grover's
algorithm succeeds with the surprisingly small amount of roughly three
i--bits, asymptotically independent of the number of qubits.  

Recently it was shown that the QFT itself on a wide variety of input states
(with efficient classical description) can be efficiently simulated
on a classical computer, as the amount of entanglement remains
logarithmically bounded \cite{Aharonov06,Yoran06,Browne06}.  As the QFT
taken by itself creates exponential 
interference, it follows that 
an exponential amount of interference alone
does not prevent an efficient classical simulation. This is in fact
obvious already from the simple (if practically useless) quantum algorithm
which consists of applying 
a Hadamard gate once on each qubit and then
measuring all qubits. By 
definition, this algorithm uses 
exponential interference  ($\cI =2^n-1$ for $n$ qubits). When applied to an
arbitrary 
computational basis state, one gets any output between $0$ and $2^n-1$ with
equal probability, $p=1/2^n$. According to Jozsa and Linden's result
\cite{Jozsa03} this algorithm cannot provide any speed-up over its classical
counterpart (as it creates zero entanglement), and indeed, it can evidently be
efficiently simulated with a simple stochastic algorithm that spits out a
random number between $0$ and $2^n-1$ with equal probability, which can be
done by choosing each bit randomly and independently equal to 0 or 1 with
probability 1/2.  We note that the same phenomenon exists also for
entanglement: 
the state $(|000...000\rangle+|111...111\rangle)/\sqrt{2}$ 
has a lot of entanglement, but the corresponding probability distribution
can be efficiently simulated classically with two registers. 

Thus the precise nature of the relationship between interference 
and the power of quantum computation is not yet fully
understood. Surprisingly, there are
tasks in quantum  
information processing which do not require interference in order to
give better than classical performance, as was shown in \cite{Lyakhov07} for
quantum state transfer through spin chains. 
In order to shed light on this question, 
in this paper we study  both Grover's and Shor's algorithms in
presence of various errors. We analyze
to what extent the interference in a 
quantum algorithm changes when the algorithm is subjected to errors, and to
what extent these changes 
reflect a degradation of the performance of the algorithm. We will
investigate this question for
systematic and random, unitary or non-unitary errors, and look
at the ``potentially available interference'' as well as the ``actually used
interference'', where the former means the interference in the entire
algorithm, the latter the interference in the part of the algorithm after
the application of the initial Hadamard gates \cite{Braun06}.

\section{Grover's and Shor's algorithms and the interference measure}\label{algos}
As we will use Grover's and Shor's algorithms throughout the paper we first
review shortly their main components. Grover's algorithm $U_G$
\cite{Grover97} finds a single marked item $a$ in an 
unstructured database 
of $N$ items in $O(\sqrt{N})$ quantum operations,
to be compared with $O(N)$ operations for the classical algorithm.
The algorithm starts on a system of $n$
qubits (Hilbert space of dimension $N=2^n$) with the Walsh-Hadamard
transform 
$W$, which transforms the computational basis state $|0\ldots 0\rangle$ into
a uniform 
superposition of the basis states
$N^{-1/2}\sum_{x=0}^{N-1}|x \rangle$.
Then the algorithm iterates $k$ times the same
operator $U= W R_2 W R_1$, with an optimal value $k=[\pi/(4\theta)]$ (where
$[\ldots]$ means the integer part) and
$\sin^2\theta=1/N$ \cite{Boyer96},
i.e. $U_G=(W R_2 W R_1)^kW$.   
The oracle $R_1$ multiplies the amplitude
of the marked item $\alpha$ with a factor $(-1)$, and keeps the other
amplitudes unchanged. 
The operator $R_2$ multiplies the
amplitude of the state $|0\ldots 0\rangle$ with a factor $(-1)$,
keeping the others unchanged. \\

Shor's algorithm \cite{Shor94} allows
the explicit decomposition of a large integer number
$R$  into prime factors in a polynomial number of operations. 
The algorithm starts by applying $2L$ Hadamard gates 
to a register of size $2L$ 
where $L=[\log_2 R]+1$, in order to 
create an equal superposition of all computational basis states
$N^{-1/2}\sum_{t=0}^{N-1}|x \rangle$ where $N=2^{2L}$.
Then the values of the function $f(x)=a^x \; (\mbox{mod} \;R)\;$, where $a$
is a randomly chosen integer with $0<a<R$,
are built on a second register of size $L$ 
to yield the state
$N^{-1/2}\sum_{t=0}^{N-1}|x \rangle|f(x)\rangle$.
The last quantum operation 
consists in a Quantum Fourier Transform (QFT) on the first
register only which allows one to find
the period of the function $f$, from which a factor 
of $R$ can be found with sufficiently high probability.
In the numerical simulations, we didn't take into account the workspace
qubits which are necessary to perform the modular exponentiation, 
but are not used elsewhere.  It is a reasonable simplification in
our case, since during this phase of modular exponentiation interference
is not modified, as the whole process is effectively a permutation
of states in Hilbert space \cite{Braun06}.
In order to study numerically the effect of errors for different
system sizes, we performed simulations for $n=12$ qubits, which corresponds,
respectively, to factorization of $R=15$ (with $a=7$), and also for $n=9$ and $n=6$.  
The cases 
$n=9$ and $n=6$ correspond to order-finding for $R=7$ (with
$a=3$) and $R=3$ (with $a=2$) and
do not exactly correspond to an actual factorization, although the 
algorithmic operations are the same, and a period is found at the end.
In the case $n=9$ the period found does not divide the dimension of the
Hilbert space, so the final wave function is not any more a superposition
of equally spaced $\delta$-peaks, but is composed of broader peaks.  
This enables to reach the more usual regime of the Shor algorithm, where
in general the period is not a power of two (although it does not happen
for $R=15$).  In one case the result was different enough to warrant the 
display of the corresponding curve for a different value of $a$
($a=6$) for which the period is a power of two (see end of Subsection
III.C).

The interference measure for a propagator $P$ of a density matrix $\rho$
($\rho_{ij}'=\sum_{k,l}P_{ij,kl}\rho_{kl}$)
derived in \cite{Braun06} is defined as
\begin{equation} \label{IM}
\cI(P)=\sum_{i,k,l}|P_{ii,kl}|^2-\sum_{i,k}|P_{ii,kk}|^2\,,
\end{equation}
where $P_{ij,kl}$ are the matrix elements of the propagator in the
computational basis $\{|k\rangle\}$ ($k=0,\ldots 2^n-1$), and
$\rho_{kl}=\langle k|\rho|l\rangle$. While the interference measure is
certainly not unique, it quantifies the two basic properties of
interference: the coherence of the propagation, and the ``equipartition'' of
the output states, i.e.~to what extent the computational basis states
are fanned out during propagation. Indeed, the second term in (\ref{IM}) can
be understood as a sum over matrix elements of a classical stochastic map
(the map which propagates the diagonal matrix elements of the density
matrix, thus the probabilities in the computational basis). This term is
subtracted from the more general first term, where the elements $P_{ii,kl}$
of the propagator are responsible for the propagation of the coherences in
the density matrix and their contribution to the final
probabilities. Therefore, if all coherences get destroyed during
propagation (i.e.~the map is purely classical), interference vanishes. The
squares in eq.(\ref{IM}) are important, as they allow to measure the
equipartition. The number of i--bits is defined as $n_I=\log_2(\cI(P))$. One
Hadamard gate provides one i-bit of interference \cite{Braun06}.

In quantum information theory the propagation of mixed states is generally
formulated within the operator sum formalism \cite{Nielsen00}: A set of
operators $\{E_l\}$ acts on $\rho$ according to $\rho'=\sum_l E_l\rho
E^\dagger_l=P\rho$, where the Kraus operators $E_l$ 
obey $\sum_k E_k^\dagger E_k={\bf 1}$ for trace--preserving operations. The
interference measure then becomes (\ref{IM})
\begin{equation} \label{IME}
\cI=\sum_{i,k,m}\left|\sum_l(E_l)_{ik}(E_l^*)_{im}\right|^2-\sum_{i,k}\left(\sum_l\left|(E_l)_{ik}\right|^2\right)^2\,.
\end{equation}
In the case of unitary propagation presented by a matrix $U$, the
interference measure reduces to
\begin{eqnarray}
\cI(P(U))
&=&N-\sum_{i,k}|U_{ik}|^4\,.\label{cohm}
\end{eqnarray}
This form makes it obvious that the interference is bounded by $0\le
\cI(P(U))\le N-1$.  The interference measure is invariant under permutations
of the computational basis states. 

\section{Perturbed quantum algorithms}
In the following we examine the amount of interference in perturbed versions
of Grover's algorithm and Shor's algorithm. 
We will distinguish between
``potentially available'' and ``actually used'' interference
\cite{Braun06}. These names 
are motivated by the fact that both algorithms start from the single
computational basis state $|0\rangle$, such that only the first column of
the unitary matrix $U$, which represents the algorithms in the computational
basis, determines the outcomes and success probabilities. The interference
measure, however, counts the interference for all possible input states,
i.e.~the contributions from all columns in $U$ --- thus the ``potentially
available'' interference is in general much larger than what is needed in
the algorithms. The actually used interference on the other hand 
is the interference in the remainder of an algorithm after all the initial
Hadamard gates have been applied. At that point a coherent superposition of
all computational basis states has been built up (in the case of Shor's
algorithm: a coherent superposition of all computational basis states of the
first register), and therefore all the interference measured by $\cI(P(U))$ is
actually used. Another motivation to look at these two different
measures is the fact that the latter focuses on the ``non--generic'' part
of the algorithm.

\subsection{Systematic unitary errors}
We start by replacing each Hadamard gate with a perturbed gate,
parametrized with an angle $\theta$, as 
\begin{equation} \label{Ht}
H(\theta)=
\left(
\begin{array}{cc}
\cos\theta & \sin\theta\\
\sin\theta & -\cos\theta
\end{array}
\right)\,.
\end{equation}
The unperturbed Hadamard gate corresponds to $\theta=\pi/4$,
while the cases $\theta=0$ and $\theta=\pi/2$ replace the
Hadamard gates by the Pauli matrices $\sigma_z$ and $\sigma_x$,
respectively, which create no interference. 
Thus the replacement of Hadamard gates by (\ref{Ht}) amounts to 
destroying the interference produced in the course of the algorithm in
a controllable fashion. This allows us to compare the loss
of interference with the efficiency of the algorithm in a systematic
way.  We measure this efficiency through the success probability $S$. For
Grover's algorithm the natural defintion of $S$ is the probability to find
the searched state $|\alpha\rangle$,
$S=\tr(\rho_f|\alpha\rangle\langle\alpha|)$, 
where the density matrix $\rho_f$ describes the final state at the
end of the computation, which may be a mixed state if decoherence strikes
during the calculation (see section \ref{sec.dec}). For
Shor's algorithm there are in general many ``good'' final states which allow
to compute the period of the function $f$, and it is therefore more
appropriate to define $S$ through the loss of probability on these ``good''
final states compared to the unperturbed algorithm. Thus, if 
$\sum_i \psi_i |i\rangle$ is the final state of the unperturbed algorithm
and $\sum_i \psi^{err}_i |i\rangle$, we define $S$ for Shor's algorithm as
\begin{equation} \label{sucprob}
S=1-\sum_i \big||\psi_i|^2-|\psi^{err}_i|^2\big|/2\,.
\end{equation}

\begin{figure}
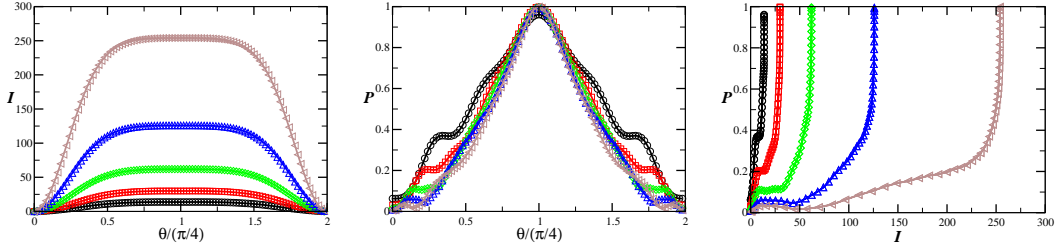

\epsfig{file=Ioft_SE_PA.eps,width=3.2cm,angle=270}\hspace{0.3cm}
\epsfig{file=Poft_SE_PA.eps,width=3.2cm,angle=270}\hspace{0.3cm}
\epsfig{file=PofI_SE_PA.eps,width=3.2cm,angle=270}
\caption{(Color online) Potentially available interference in the Grover
  algorithm with systematic 
  unitary errors in the Hadamard gates, parametrized by the angle $\theta$,
  eq. (\ref{Ht}) (left); success probability $S$ of the algorithm (middle); and
  success probability as function of interference (right). Black circles
  mean $n=4$, red squares $n=5$, green diamonds $n=6$, blue
  triangles up $n=7$, 
  gray triangles left $n=8$. All curves are averaged over all values of the
  searched item $\alpha$. 
  }\label{fig.SEPA}        
\end{figure}

Figure
\ref{fig.SEPA} shows the dependence of the potentially available
interference and of  the success probability $S$ of Grover's algorithm on
$\theta$, as well as the success probability as function of the
interference.  All curves are averages over
all values of $\alpha$, $\alpha=0,\ldots,2^n-1$. Both interference and success
probability peak at $\theta=\pi/4$. For a small number of qubits,
$S(\theta)$ shows some additional modulation in the wings of the $n$ curve,
which are pushed further and further out for increasing $n$. The broad
maxima of $\cI(\theta)$ lead to steep increases of $S(\cI)$ close to the
maximum possible value for the interference $\cI=2^n-1$. At $\theta=0$ or
$\theta=\pi/2$, the interference vanishes, as in that case the algorithm
degenerates to a combination of permutations and phase shifts, so that no
two computational states get superposed. 
Figure \ref{fig.SEPA} shows that for this example an exponential amount of
interference is necessary even in order to obtain a success probability of
the order $1/2$, and by squeezing out a small additional amount of
interference, $S$ can be boosted to its optimal value.

\begin{figure}
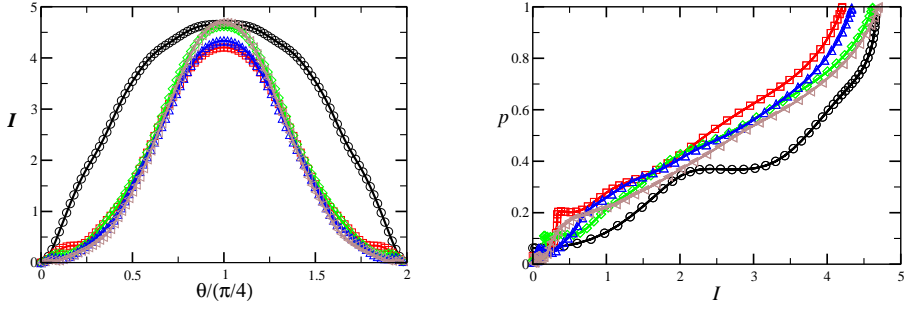

\epsfig{file=Ioft_SE_AU.eps,width=4cm,angle=270}\hspace{1cm}
\epsfig{file=PofI_SE_AU.eps,width=4cm,angle=270}
\caption{(Color online) Same as Fig.~\ref{fig.SEPA}, but for actually used
  interference. }\label{fig.SEAU}        
\end{figure}

The actually used interference gives a similar picture: For $\theta=0$
or $\theta=\pi/2$ the interference vanishes, and interference reaches its
maximum value $\cI\simeq 4$ for $\theta=\pi/4$. As the success probability
remains unchanged whether we calculate the interference for the entire
algorithm or only after the initial Hadamard gates, we find again that the
success probability increases with increasing interference 
(see Fig.\ref{fig.SEAU}).

\
\begin{figure}
\epsfig{file=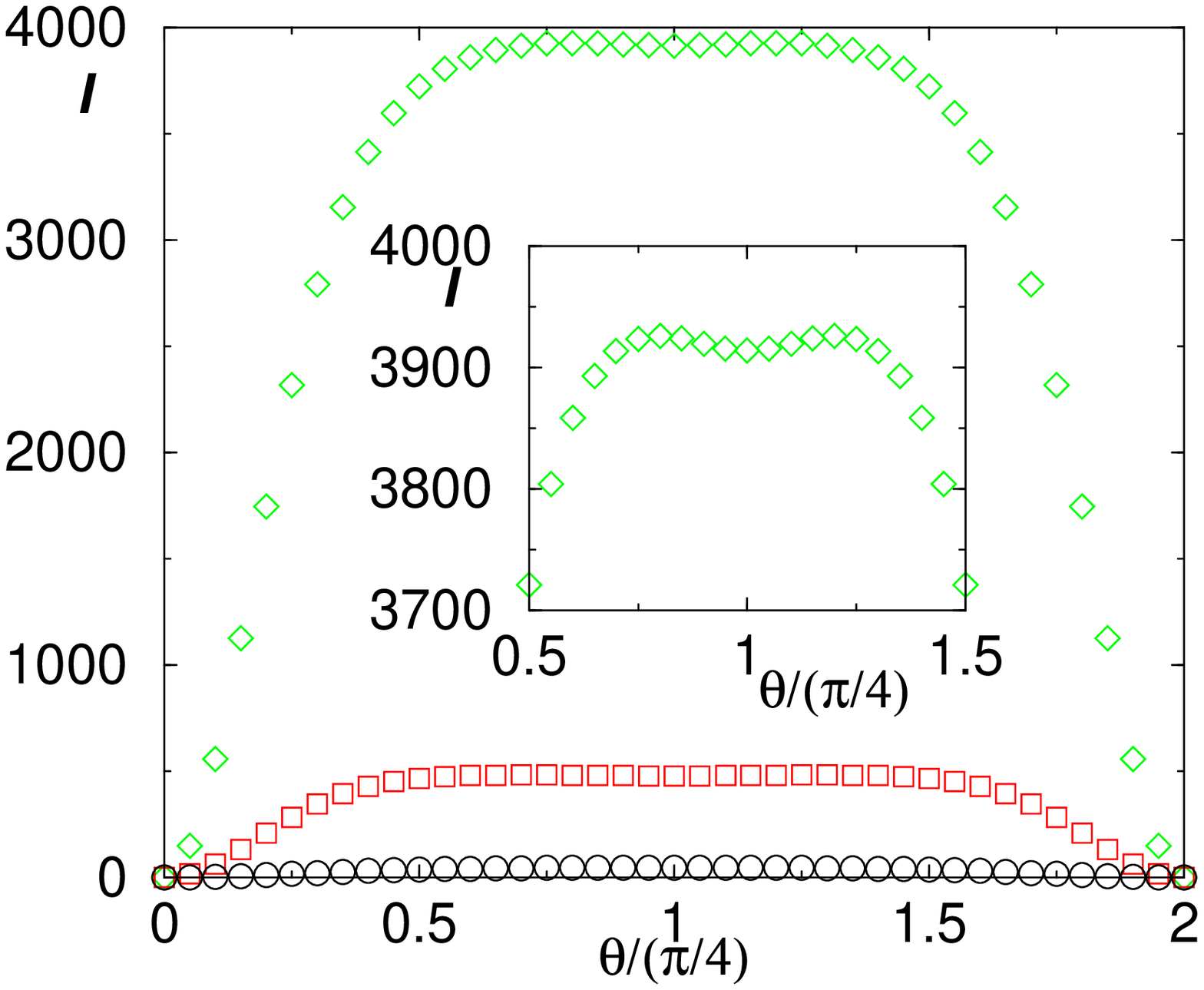,width=5cm,angle=0}\hspace{0.3cm}
\epsfig{file=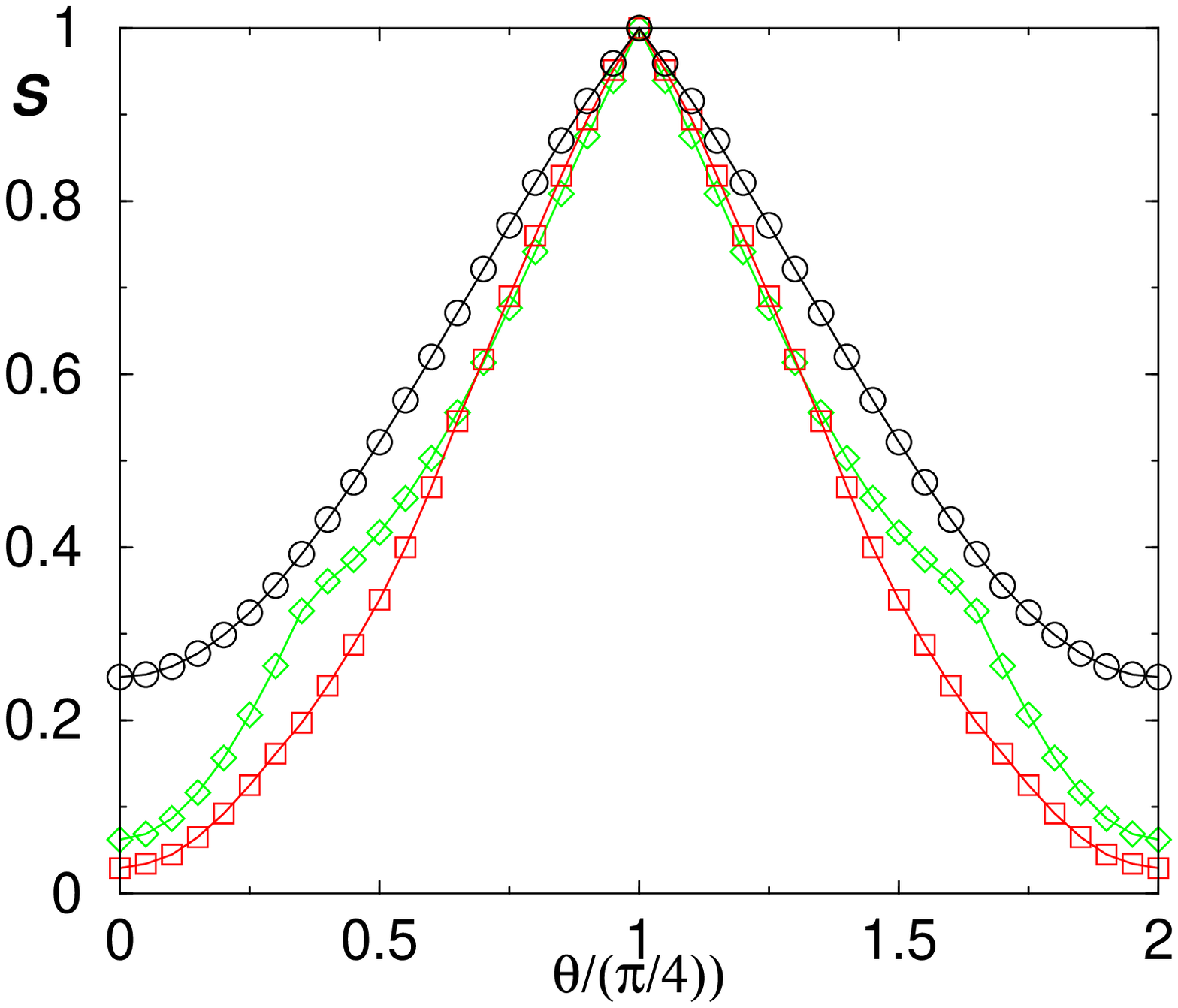,width=5cm,angle=0}\hspace{0.3cm}
\epsfig{file=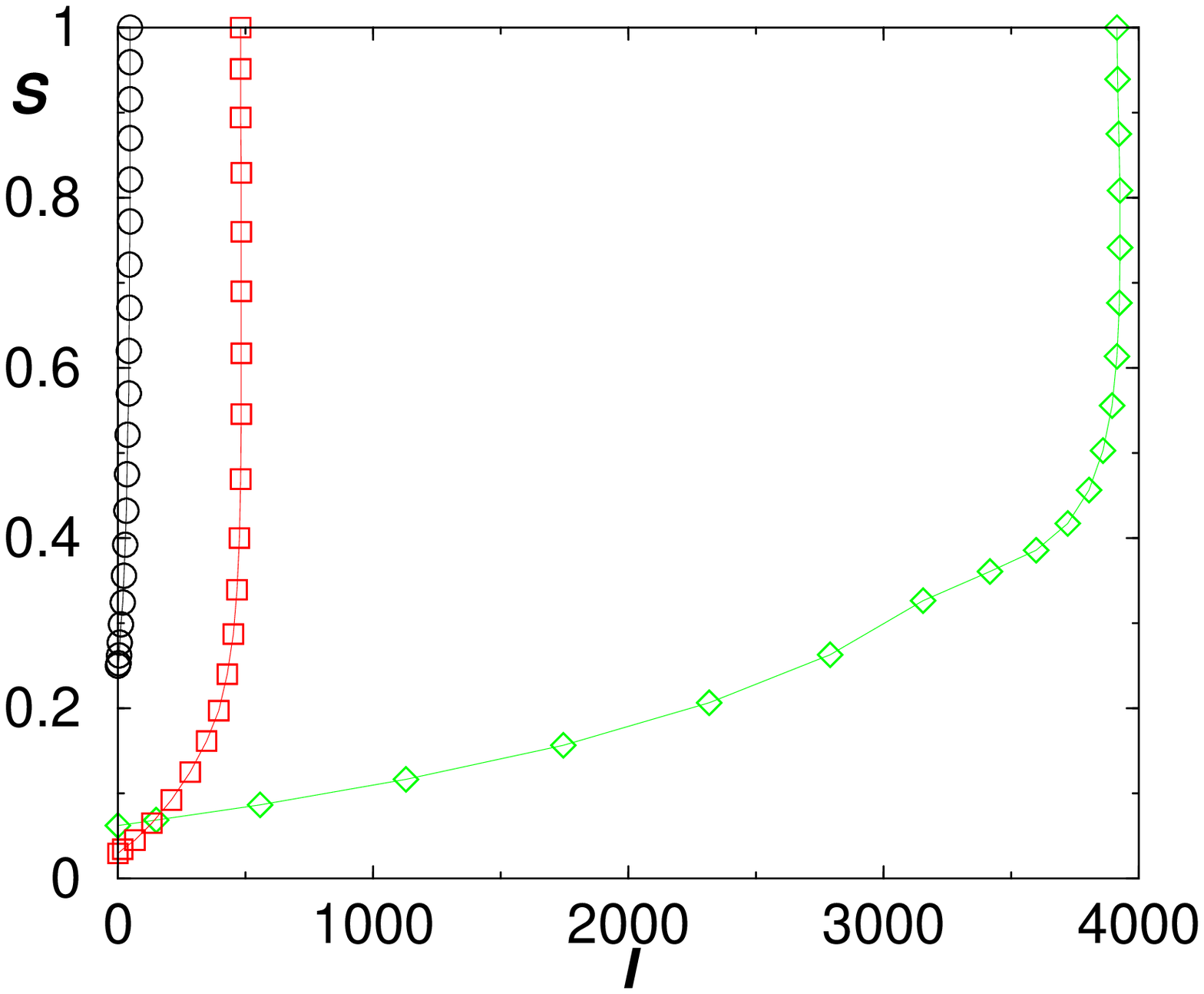,width=5cm,angle=0}
\caption{(Color online) Potentially available interference in the Shor
  algorithm with systematic 
  unitary errors in the Hadamard gates, parametrized by the angle $\theta$,
  eq. (\ref{Ht}) (left); success probability $S$ of the algorithm (middle); and
  success probability as function of interference (right). Black circles mean
  $n=6$ ($f(x)=2^x \; (\mbox{mod} \;3)\;$), red squares $n=9$ ($f(x)=3^x \; (\mbox{mod} \;7)\;$), green diamonds $n=12$ 
($f(x)=7^x \; (\mbox{mod} \;15)\;$).  Inset on the left is a close-up of the 
case $n=12$ close to the maximum.
  }\label{fig.shor1}        
\end{figure}

Figure
\ref{fig.shor1} shows the potentially available
interference and  the success probability 
for Shor's algorithm.
Again, both interference and success
probability peak at $\theta=\pi/4$. 
The additional modulation in the wings of the curve for $S(\theta)$
for a small number of qubits is much less pronounced 
than for Grover's algorithm, 
but the main fact remains that the broad
maximum of $\cI(\theta)$ corresponds to a sharp peak for $S(\theta)$,
leading to the same steep increase of $S(\cI)$ close to the
maximum possible value of the interference.
However, the exact algorithm does not lead to the maximum 
possible amount of interference. Close to $\theta=\pi/4$, the interference
slightly increases while the success  
probability goes down, indicating that some interference which is ``useless'' 
in terms of the algorithm efficiency is generated.  When $\theta$ increases
and the interference is reduced by a large amount, the algorithm
 has a low success probability.  Even though the success probability
globally 
goes down with an increasing number of qubits, it is not the case   
for all $\theta$ values.  This can be attributed to the fact that
the three curves on the figure do not exactly describe the same problem for 
different numbers of qubits, but are instances of order-finding for different
values of the number $R$. Thus when 
different values of the number of qubits are used, the precise
problem investigated depends on the number theoretical properties
of the integers chosen, which can be different.  On the contrary,
Grover's algorithm run on different numbers of qubits is essentially
the same problem run on a computer of different size.

\begin{figure}
\epsfig{file=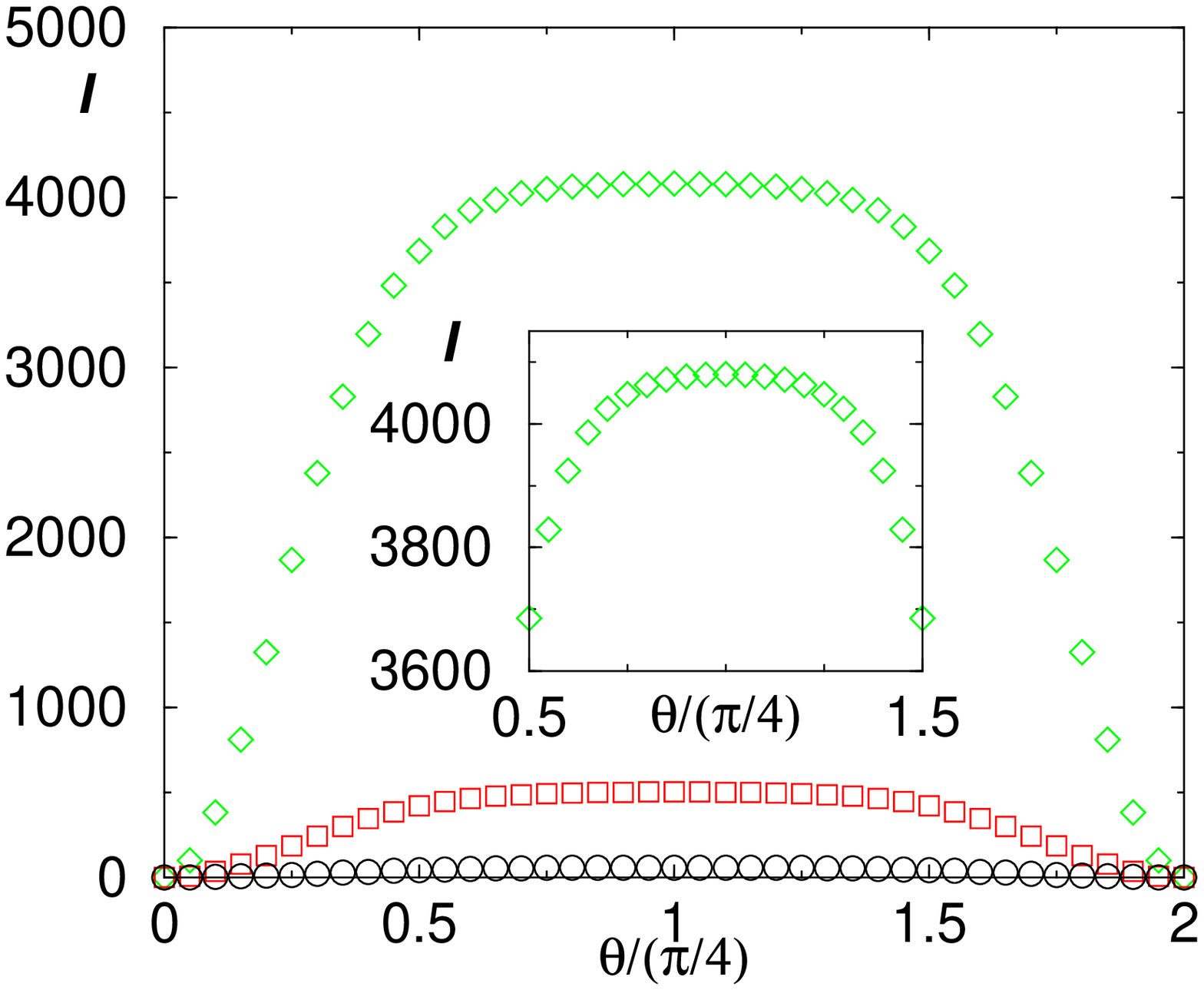,width=6cm,angle=0}\hspace{0.3cm}
\epsfig{file=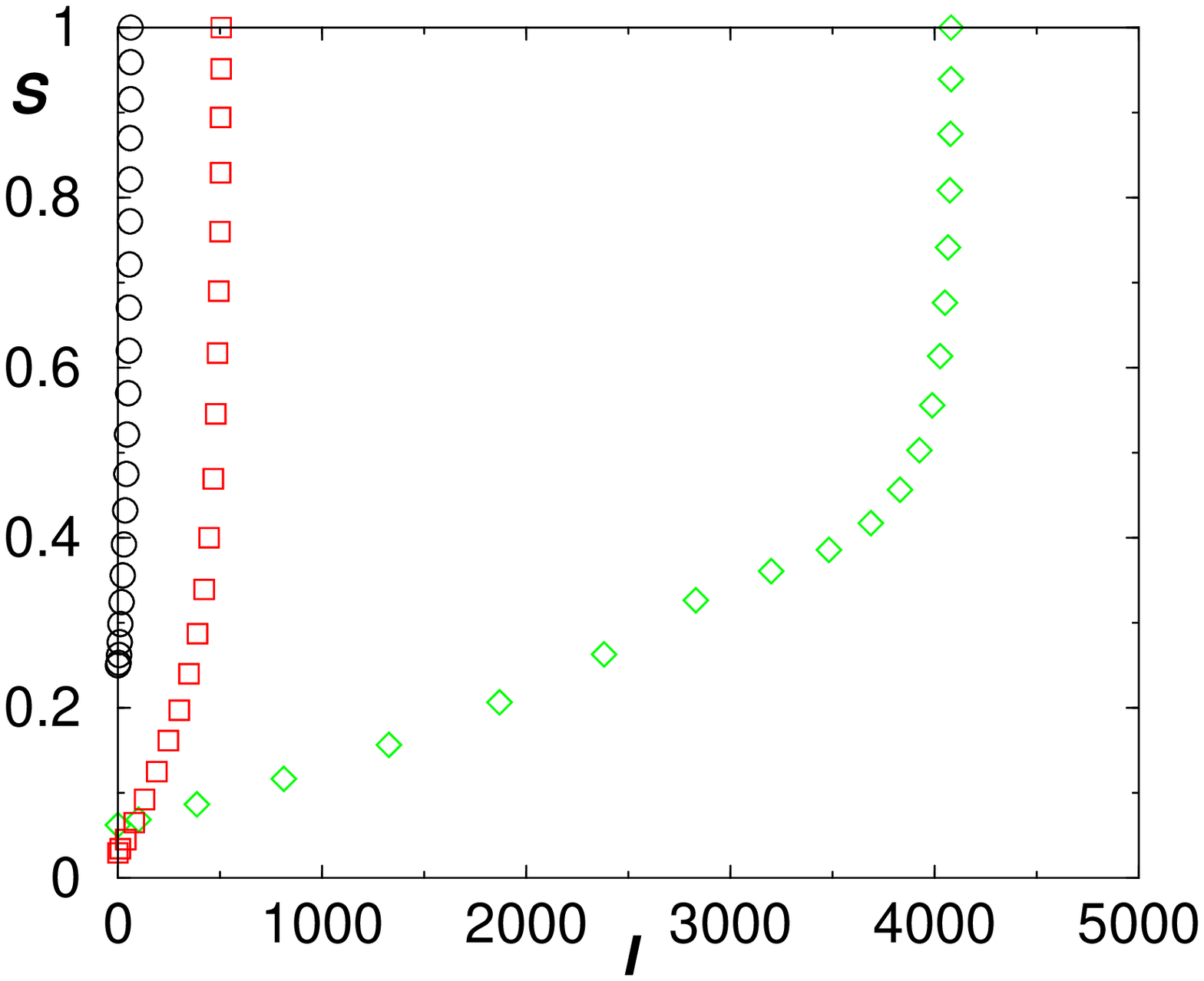,width=6cm,angle=0}
\caption{(Color online) Same as Fig.~\ref{fig.shor1}, but for actually used
  interference. }\label{fig.shor2}        
\end{figure}

In \cite{Braun06}, it was pointed out that Grover's and Shor's
algorithm use a very different amount of actually used interference.
Indeed, for Grover's algorithm it remains bounded for all values
of the number of qubits $n$ , while for Shor's algorithm it grows
exponentially with $n$.  This may be related to the fact that Shor's
algorithm is exponentially faster than all known classical
algorithms, while Grover's is only quadratically faster. 
In Figure \ref{fig.shor2} the 
actually used interference is plotted for Shor's algorithm, 
showing that for $\theta=\pi/4$ it reaches its maximal value which grows
exponentially with the number of qubits.  In this case, any decrease in
the interference  corresponds to a decrease of the success probability.

\subsection{Random unitary errors} 
Let us now consider what happens if we replace each Hadamard gate with a
gate given by eq.(\ref{Ht}), where each $\theta$ is chosen randomly,
uniformly and independently from all other gates  in an interval
$\pi/4-\epsilon/2,\pi/4+\epsilon/2$.
\begin{figure}
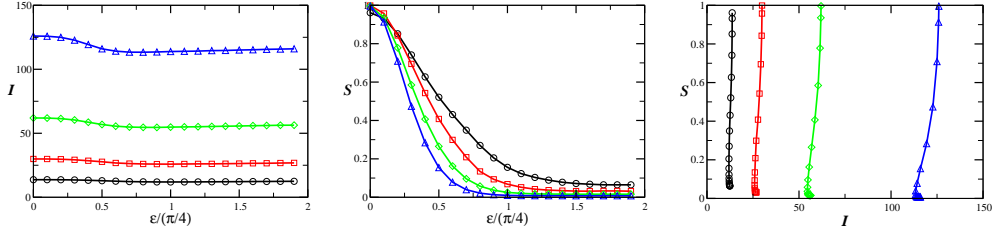

\epsfig{file=Iofe_RE_PA.eps,width=3cm,angle=270}\hspace{0.3cm}
\epsfig{file=Pofe_RE_PA.eps,width=3cm,angle=270}\hspace{0.3cm}
\epsfig{file=PofI_RE_PA.eps,width=3cm,angle=270}
\caption{(Color online) Potentially available interference in the Grover
  algorithm with random
  unitary errors in the Hadamard gates, parametrized by the interval
  $\epsilon$, 
  eq.(\ref{Ht}) (left); success probability $S$ of the algorithm (middle); and
  success probability as function of interference (right) for $n=4$ to
  $n=7$. Same symbols as 
  in Fig.~\ref{fig.SEPA}. 
  }\label{fig.REPA}        
\end{figure}

Figure \ref{fig.REPA} shows the interference and success probability of 
Grover's algorithm as function of $\epsilon$. All curves are averaged over all
possible values of $\alpha$ as well as over $n_r$ random realizations of the
algorithm ($n_r=1000$ for $n=4$, $n_r=100$ for $n=5,\ldots,7$). Again, the 
maximum amount of 
interference is obtained for the unperturbed algorithm, $\epsilon=0$, but
the maximum is not very prominent. It is followed by a shallow minimum close to
$\epsilon=\pi/4$, which gets shifted to smaller values for increasing
$n$. Altogether, the interference is little affected by the random unitary
errors. This can be understood from the fact that the unitary matrix $U$
representing the algorithm is already almost full in the unperturbed
algorithm \cite{Braun06}, with the exception of the first column, which
propagates the initial state $|0\rangle$, and presents a strong peak on the
searched item. Randomly replacing the Hadamard gates by $H(\theta)$
increases the equipartition in the first column, as is witnessed by the
decay of $S$, but reduces on average the equipartition in the other
columns, leading to a slight overall decrease of interference close to
$\epsilon=0$. This means again that a very large amount of interference is
necessary in order to get even a modest performance of the algorithm, and
a very steep increase of the success probability occurs when interference is
boosted to its maximum value.  

\begin{figure}
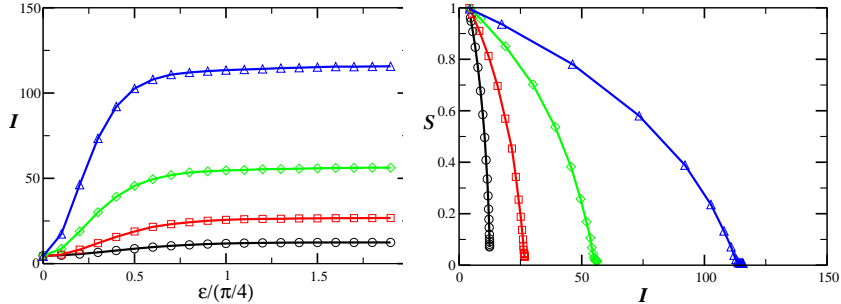

\epsfig{file=Iofe_RE_AU.eps,width=4cm,angle=270}
\epsfig{file=PofI_RE_AU.eps,width=4cm,angle=270}
\caption{(Color online) Actually used interference in the Grover
  algorithm with random unitary errors in the Hadamard gates, parametrized
  by the interval 
  $\epsilon$,  eq.(\ref{Ht}) (left), and 
  success probability $S$ as function of interference (right). Same symbols as
  in Fig.~\ref{fig.SEPA}. The success probability as function of $\epsilon$
  is the same as in Fig.~\ref{fig.REPA}. 
  }\label{fig.REAU}        
\end{figure}

The situation for the actually used interference is quite different, as 
shows Fig.~\ref{fig.REAU}. For $\epsilon=0$, we have an interference $\cI\sim 4$
at the end of the algorithm (after the first diffusion gate it reaches its
maximum value of $\cI\sim 8-24/N$ and then oscillates and decays with each
subsequent diffusion gate to the final value $\cI\sim 4$
\cite{Braun06}). Thus, the unperturbed algorithm leads to remarkably low
equipartition in the entire matrix $U$, a highly unlikely situation for any
random matrix. Indeed it was shown in \cite{Arnaud07} that a random unitary
$N\times N$ matrix
drawn from the circular unitary ensemble (CUE) gives, with almost certainty, an
interference $\cI\sim N-2$. Thus, it is not surprising that with growing
$\epsilon$, $\cI$ rapidly increases to a value $\cI\sim N$. As the success
probability decreases with $\epsilon$, this leads to the 
counter--intuitive situation that the success probability decays
with increasing actually used interference.

\begin{figure}
\epsfig{file=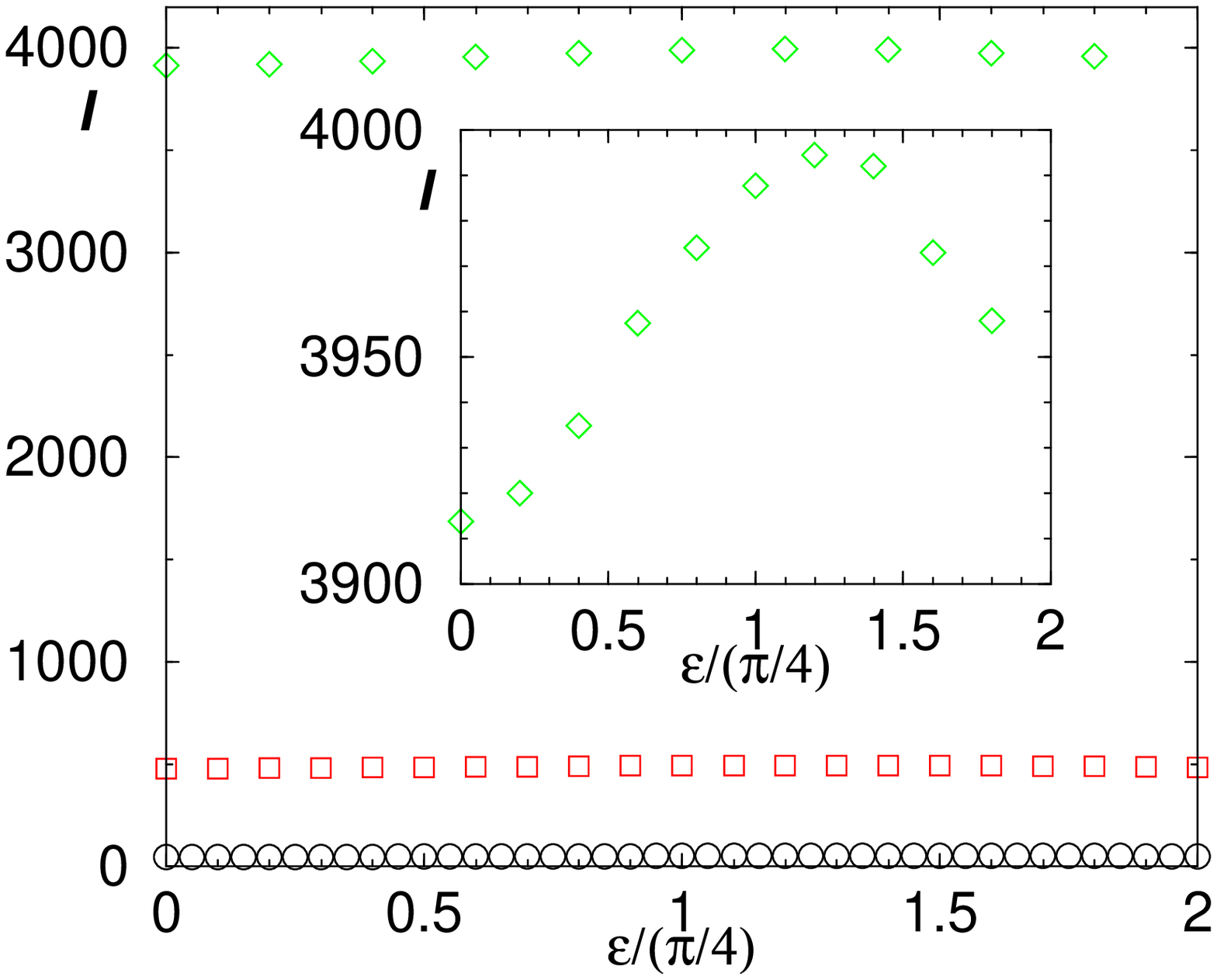,width=5cm,angle=0}\hspace{0.3cm}
\epsfig{file=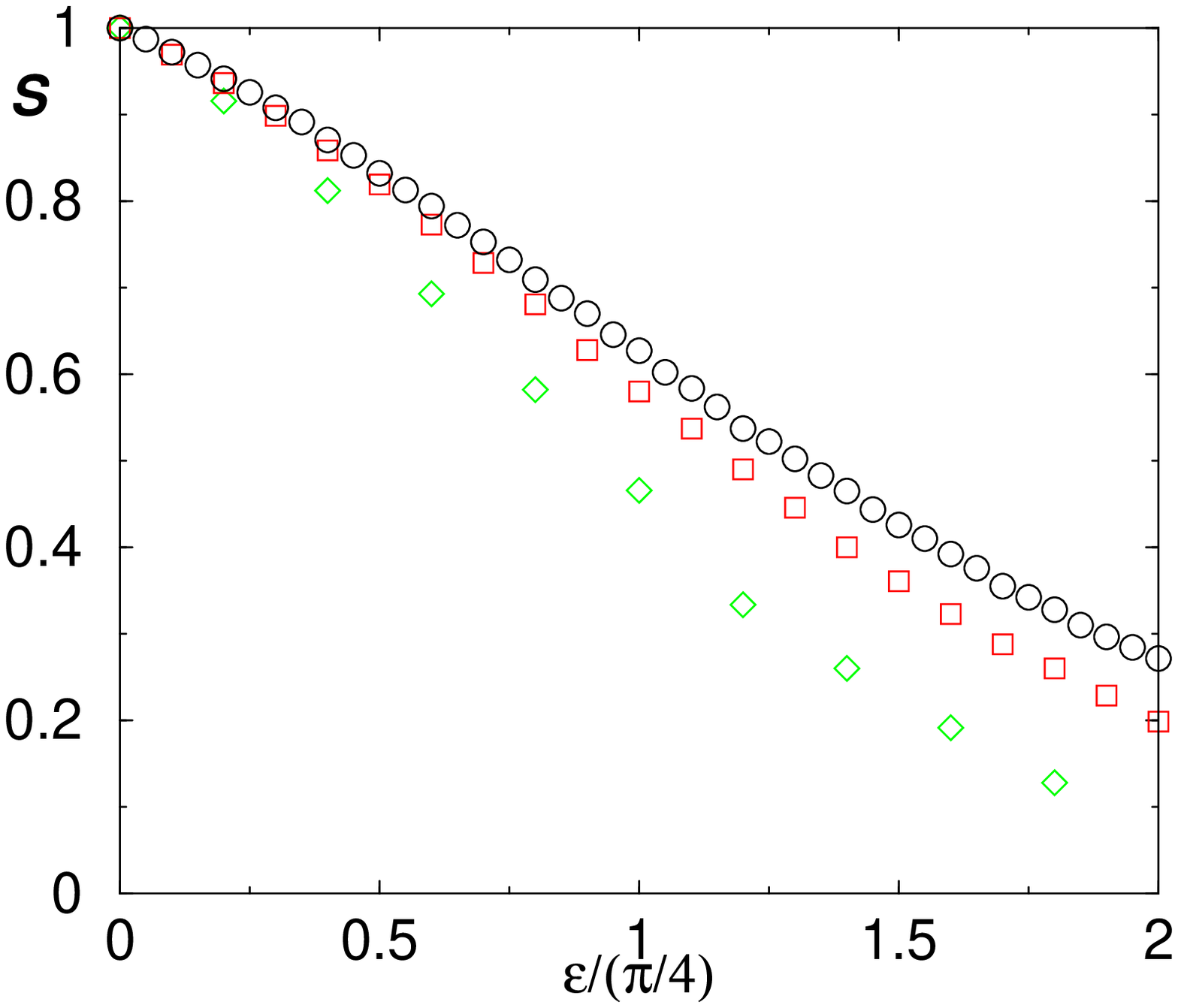,width=5cm,angle=0}\hspace{0.3cm}
\epsfig{file=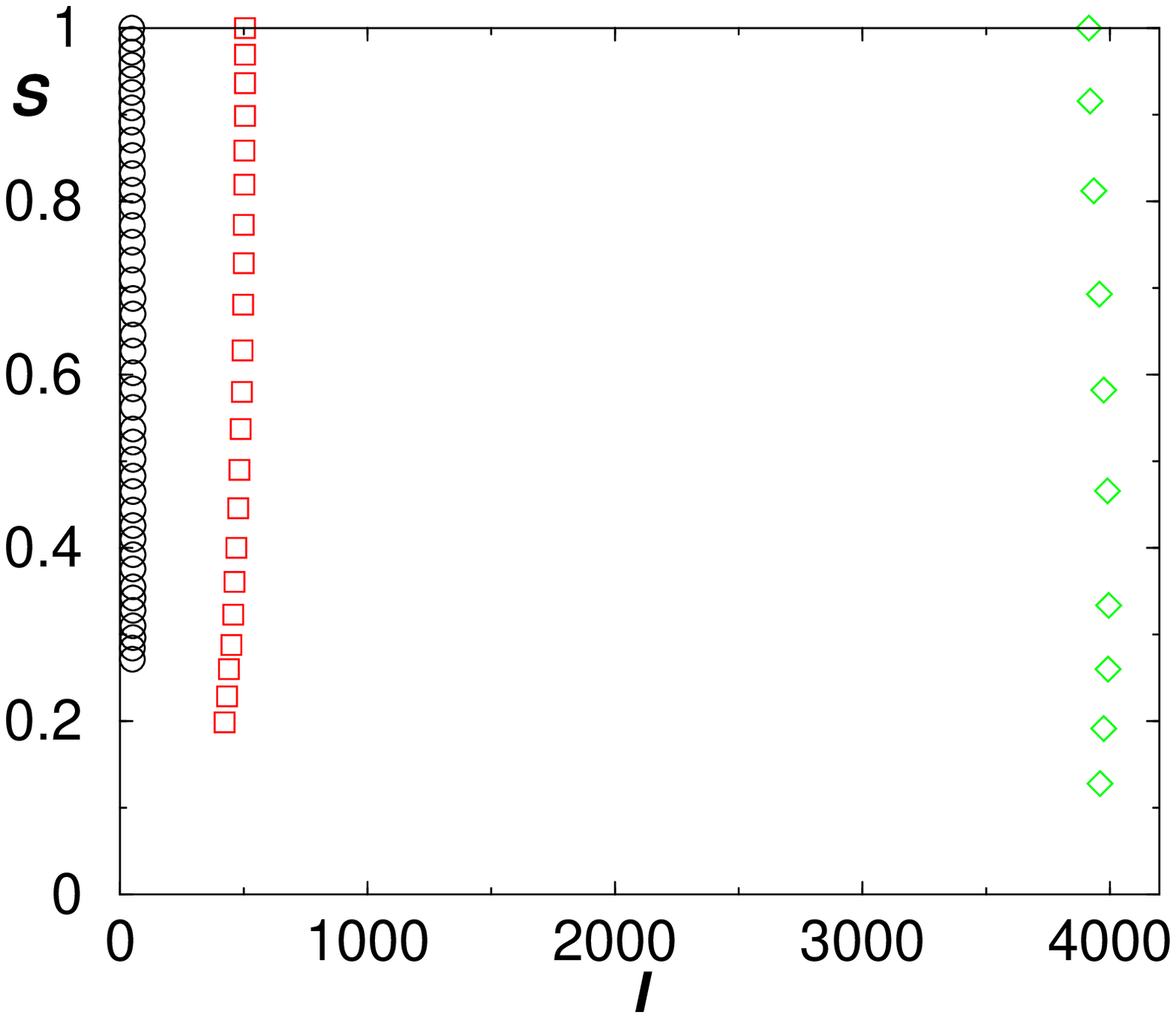,width=5cm,angle=0}
\caption{(Color online) Potentially available interference in Shor's
  algorithm with random
  unitary errors in the Hadamard gates, parametrized by the interval
  $\epsilon$, 
  eq.(\ref{Ht}) (left); success probability of the algorithm (middle); and
  success probability as function of interference (right). Symbols as in
  Fig.~\ref{fig.shor1}.  Inset shows a close-up of the curve for $n=12$. 
  }\label{fig.shor3}        
\end{figure}

\begin{figure}
\epsfig{file=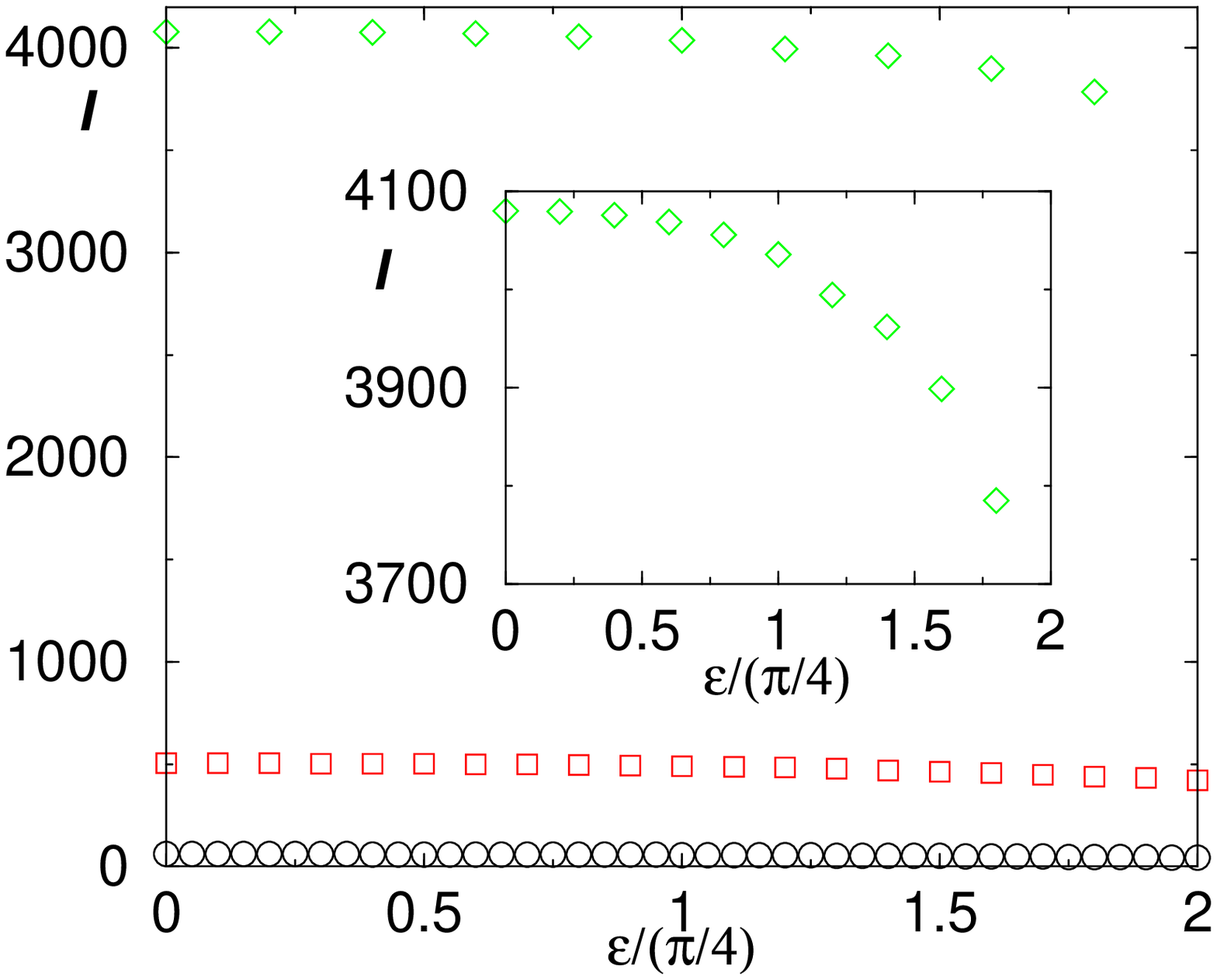,width=6cm,angle=0}
\epsfig{file=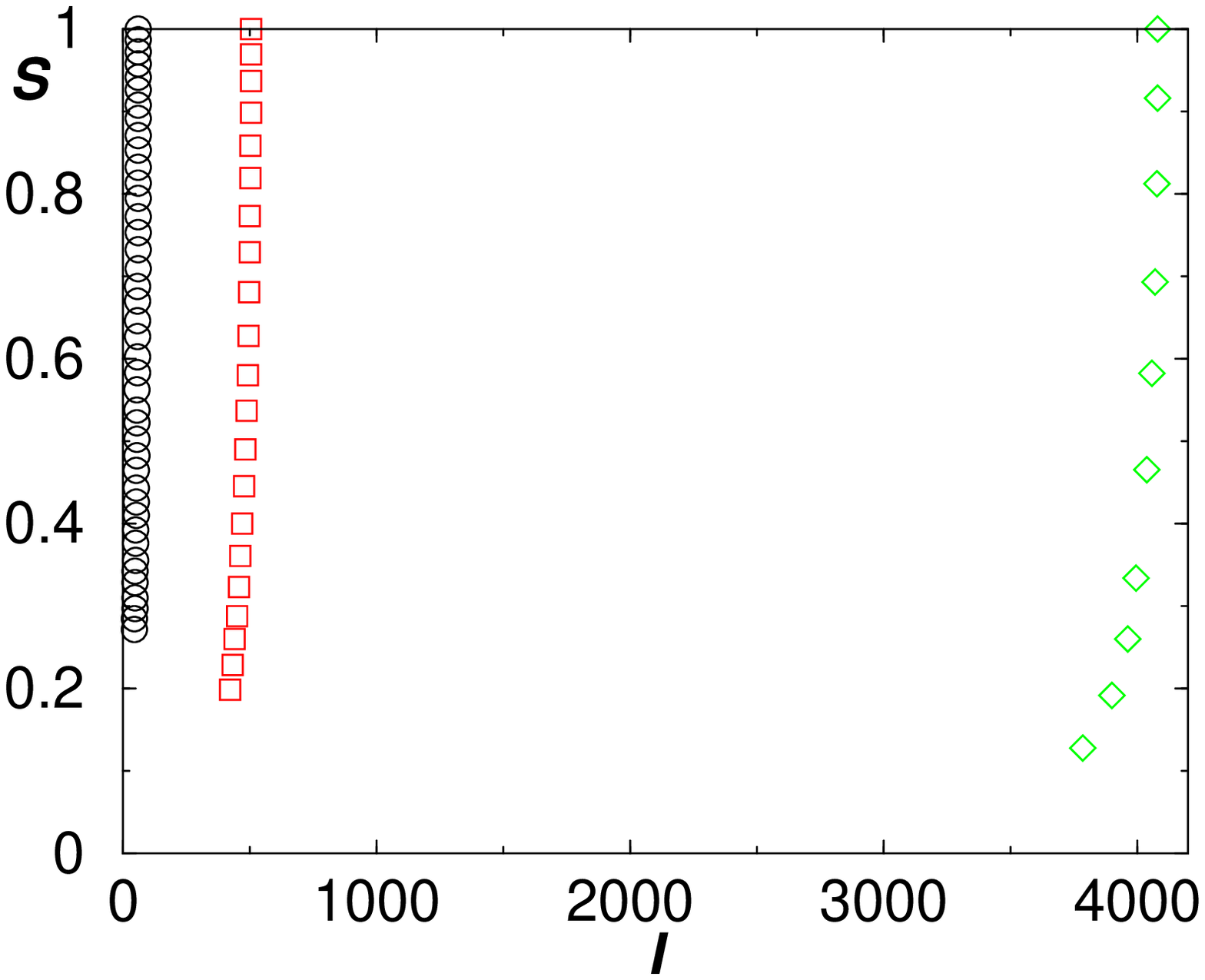,width=6cm,angle=0}
\caption{(Color online) Actually used interference in Shor's
  algorithm with random unitary errors in the Hadamard gates, parametrized
  by the interval 
  $\epsilon$,  eq.(\ref{Ht}) (left), and 
  success probability as function of interference (right). Same symbols as
  in Fig.~\ref{fig.shor1}. 
Inset shows a close-up of the curve for $n=12$.
The success probability as function of $\epsilon$
  is the same as in Fig.~\ref{fig.shor3}.
  }\label{fig.shor4}        
\end{figure}

Figures \ref{fig.shor3},\ref{fig.shor4} display the effect of random unitary
errors on Shor's algorithm with the number of random realizations $n_r=5000$
($n=6$), $n_r=1000$ ($n=9$), and $n_r=100$ ($n=12$). Besides changing the
Hadamard gates, random phases with the same distribution were added to the
two--qubit gates in the 
quantum Fourier transform. This way in both algorithms, all Fourier
transforms and Walsh-Hadamard transforms are randomized in a comparable
way. Figure \ref{fig.shor3} 
shows that the
potentially available interference oscillates slowly as a function of
$\epsilon$, on a scale which seems independent of the number of qubits
and also larger than for Grover's algorithm. The situation is
similar to the one in Fig.~\ref{fig.REAU} (although 
Fig.~\ref{fig.REAU} deals with actually used interference),
 since interference 
increases for small values of $\epsilon$ and reaches a maximal
value around $\epsilon=0.6-0.7$, while the success probability decreases.

The situation is different in the case of actually used interference,
shown in Fig.~\ref{fig.shor4}.  Indeed, interference starts from its maximum
possible value and decreases with increasing $\epsilon$ at the same time as
the success probability decreases.   In the same way as for
potentially available interference, 
the variation of interference is relatively small 
compared to the case of systematic errors, which were explicitly designed to 
destroy interference.  Nevertheless, Fig.~\ref{fig.shor4} shows that contrary
to the case of Grover's algorithm, interference and success probability
decrease in a correlated way.

\subsection{Decoherence}\label{sec.dec}
We finally consider a class of errors which create true decoherence. We
distinguish 
between phase flips and bit flips, and consider a (somewhat
artificial) situation, where the errors occur only during the first
Walsh--Hadamard transformation, i.e.~the sequence of Hadamard gates on all
qubits at the beginning of the algorithm in the case of Grover's algorithm,
and on all qubits of the first register of length $2L$ in the case of Shor's
algorithm.  We will assume that $n_f$
out of $n$ qubits are affected by errors, and study interference and success
probability as function of $n_f$, $n_f=1,\ldots,n$. Note that if all
Hadamard gates in the entire algorithm were prone to error, one would need
to calculate $2^{(2k+1)n}$ Kraus operators for Grover's algorithm (see
sec.\ref{algos}), each of 
which is a $2^n\times 
2^n$ matrix, which makes the numerical
calculation rapidly too costly. The former number is reduced to a more
bearable $2^{n_f}$ in our 
case. For Shor's algorithm, the number of Hadamard gates depends on the
implementation of the modular exponentiation and the calculation of the
function $f$, but grows exponentially with the number of qubits as well, if
all qubits can be affected by the decoherence process. 
Contrary to the calculations for unitary errors, in the simulation
of Grover's algorithm we restricted
ourselves to a fixed value of the searched item $\alpha$, but checked for a few
different values of $\alpha$ that the results are insensitive to the value
of $\alpha$.  

A Hadamard gate prone to errors   
is followed with probability $p$ by a bit--flip (or by a phase flip --- we
consider only one type of error at a time), and we calculate again both
the potentially available and actually used interference. In the latter
case, only the Pauli operators $\sqrt{p}\sigma_z$ and $\sqrt{p}\sigma_x$
which represent the phase flip and bit flip errors, respectively, with
probability $p$ on 
a given qubit are included in the Kraus operators, but not the initial
Hadamard gates themselves.

\begin{figure}
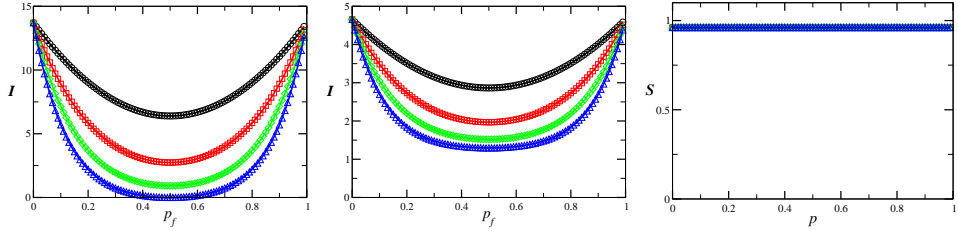

\epsfig{file=Ioft_DEC_PA_bitflip.eps,width=3cm,angle=270}\hspace{0.1cm}
\epsfig{file=Ioft_DEC_AU_bitflip.eps,width=3cm,angle=270}\hspace{0.1cm}
\epsfig{file=Poft_DEC_PA_bitflip.eps,width=3cm,angle=270}
\caption{(Color online) Potentially available interference in the Grover
  algorithm with decoherence through bit--flips during the first
  Walsh-Hadamard transformation, as function of the bit--flip probability
  $p$ after each Hadamard gate (left). Same but for actually used
  interference (middle). Success probability as function of
  $p$ (right). All curves are for $n=4, \alpha=2$; $n_f=1$ black circles,
  $n_f=2$ red 
  squares, $n_f=3$ green diamonds, $n_f=4$ blue triangles.
  }\label{fig.DECBFPA}        
\end{figure}

\begin{figure}
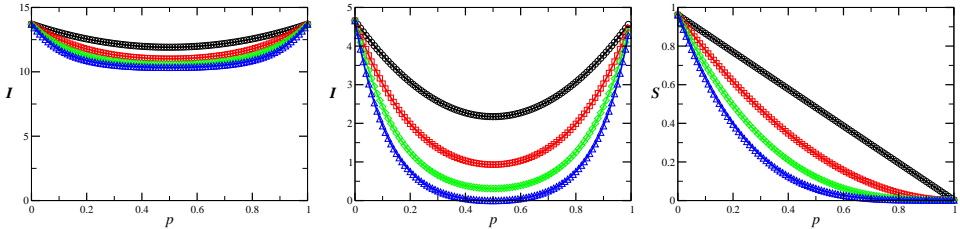

\epsfig{file=Ioft_DEC_PA_phaseflip.eps,width=3cm,angle=270}\hspace{0.1cm}
\epsfig{file=Ioft_DEC_AU_phaseflip.eps,width=3cm,angle=270}\hspace{0.1cm}
\epsfig{file=Poft_DEC_PA_phaseflip.eps,width=3cm,angle=270}
\caption{(Color online) Same as Fig.~\ref{fig.DECBFPA}, but for phase flip
  errors.
  }\label{fig.DECPFPA}        
\end{figure}

For Grover's algorithm, Fig.~\ref{fig.DECBFPA} shows the result in the
case of bit flip errors, and 
Fig.~\ref{fig.DECPFPA} for phase flip errors. 
For both types of error, the interference has maximal value for $p=0$ or
$p=1$, which 
corresponds to completely coherent propagation, and minimal value for
$p=0.5$. The minimal value decreases rapidly with the number of qubits
prone to error. The potentially available interference reaches zero in the
case of bit flip errors on all $n$ 
qubits, whereas for phase flip errors a finite value remains. The actually
used interference shows the opposite behavior. It has zero minimal value
for phase flip errors on all $n$  
qubits, whereas it remains finite for bit flip errors even with probability
$0.5$. 
Phase flip errors rapidly destroy the operability of the algorithm. The
success probability decreases linearly with $p$ for $n_f=1$ to reach
a value close to zero for $p=1$, and more and more rapidly for increasing
$n_f$. In fact, for $n=4$, $S(p=1)\simeq 0.0025$, independent of $n_f$,
which is even smaller than the classical value 1/16=0.0625. The
algorithm is completely coherent in this case, and the large amount of
interference is used in a destructive way, {\em subtracting} probability
from the searched item. 

Remarkably, bit--flip errors do not affect the success probability  
at all, such that $S(p)$ remains constant at the
optimal value, independently of the number of qubits affected.
The behavior is easily understood, as in fact the bit flip errors
leave the state obtained after applying the Hadamard gates invariant (and
in particular: pure). The interference goes down to zero, nevertheless, as
it measures coherence using superpositions of all computational states
\cite{Braun06}, and not pureness of the final state. We have therefore the
peculiar situation 
where in spite of decoherence processes one particular state remains pure
(the perfectly equipartitioned superposition of all computational basis
states), and since it is that state which is used in the algorithm, the
success probability remains unaffected. On the other hand, the interference
measure was constructed  to measure coherence by the sensitivity of final
probabilities to {\em
  relative} initial phases between the computational basis input states, and
it correctly picks up that the phase coherence {\em between} all states got
lost. Thus, in this particular situation, one can have a perfectly well
working algorithm which uses, according to our measure, zero potentially
available 
interference. 
We believe, however, that this case where
the coherence of the propagation cannot be measured by the influence of
relative phases but only through the purity of the
final state  is highly
exceptional and 
should not exclude a proof  that exponential speed--up needs
exponential interference, if one restricts attention to unitary
algorithms, or gives special attention to the exceptional single pure state
mentioned. 
It is also important to note that the actually used interference remains
finite at $p=0.5$ and $n_f=n$, and 
below the already small value $\cI\sim 4$ for the unperturbed algorithm and
thus never goes to zero for bit-flip errors.

\begin{figure}
\epsfig{file=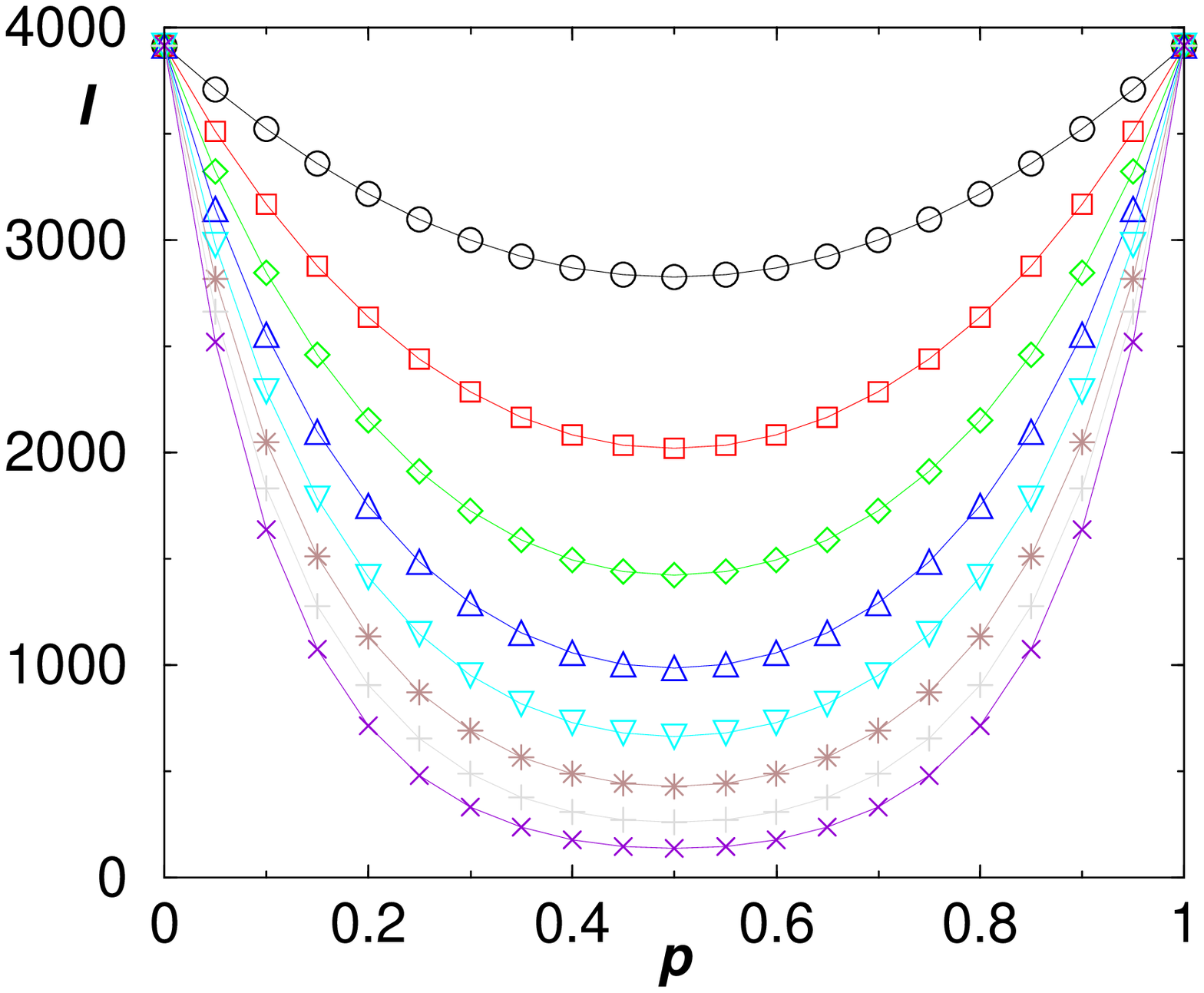,width=5cm,angle=0}
\epsfig{file=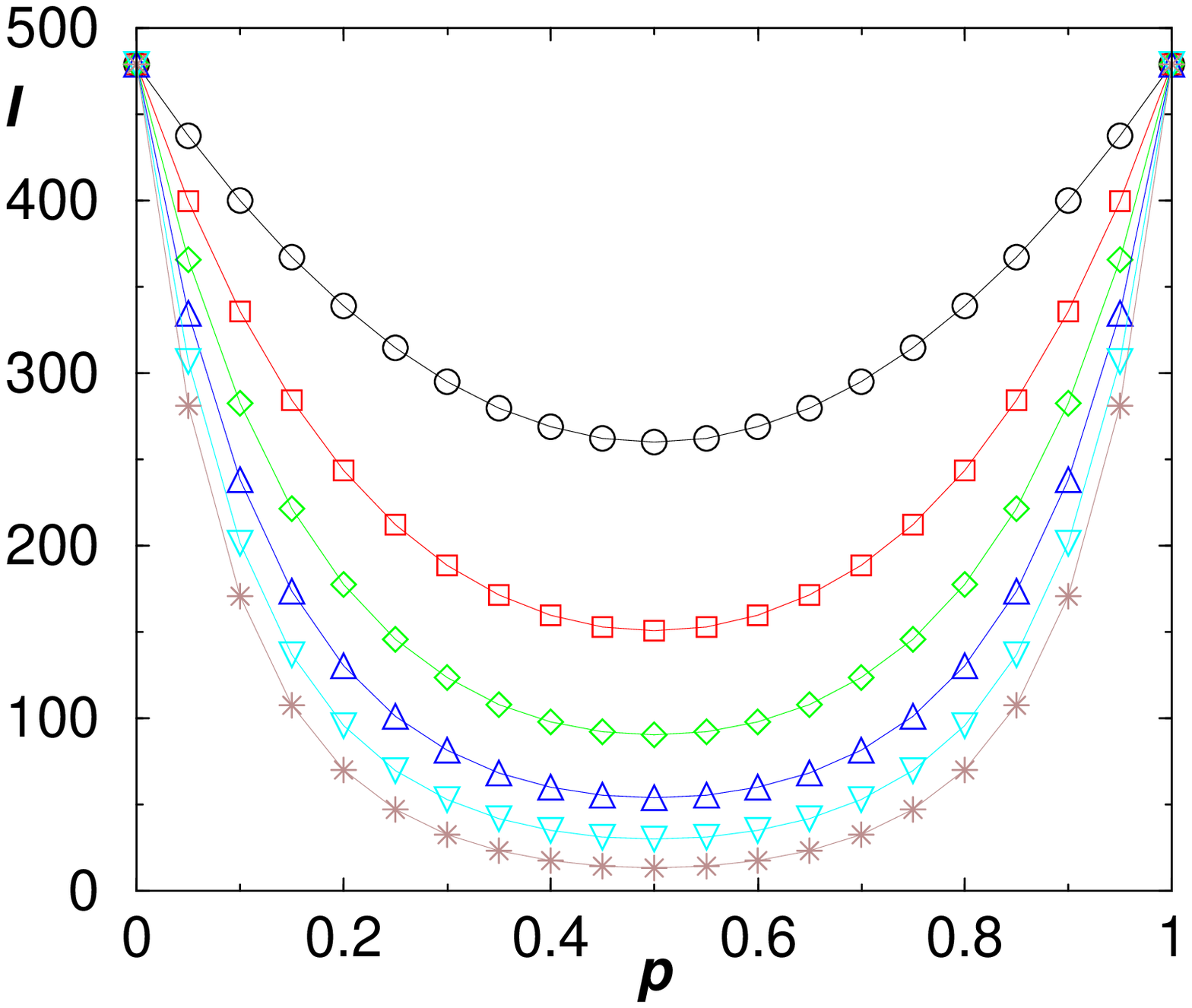,width=5cm,angle=0}
\epsfig{file=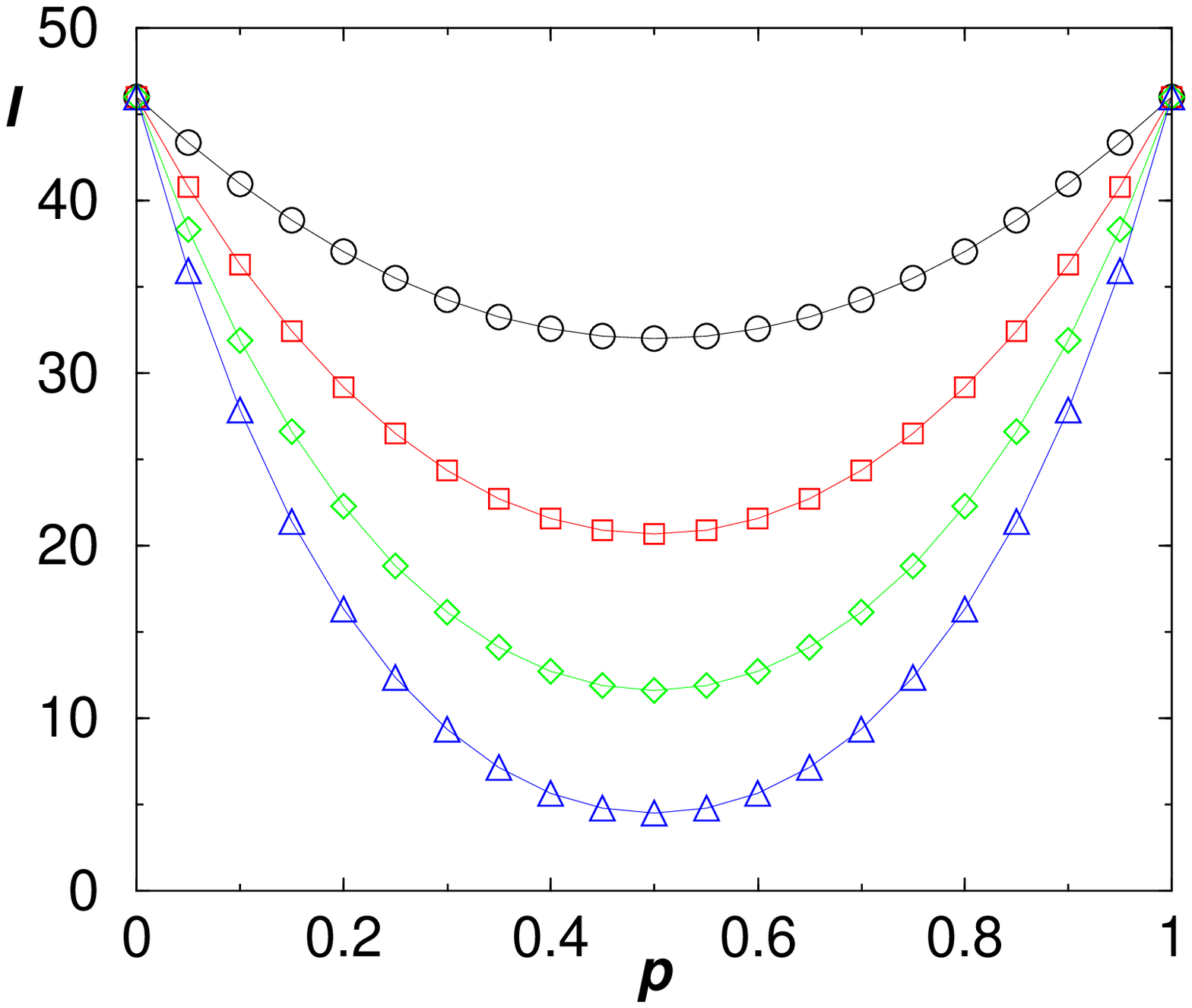,width=5cm,angle=0}
\caption{(Color online) Potentially available interference in the Shor
  algorithm with decoherence through bit--flips during the first
  Walsh-Hadamard transformation, as function of the bit--flip probability
  $p$ after each Hadamard gate, for  
$n=12$ (left), $n=9$ (center), $n=6$ (right).
The symbols are: $n_f=1$ black circles,
  $n_f=2$ red 
  squares, $n_f=3$ green diamonds, $n_f=4$ blue triangles up,
$n_f=5$ cyan triangles down, $n_f=6$ brown stars, $n_f=7$ gray
$\times$, $n_f=8$ violet $+$.  The corresponding success probability $S$
is constant equal to $p=1$ for all values of $p$ (data not shown). Here and
  in the following figures all quantities are averaged over all possible
  choices of the $n_f$ qubits in the first register.
  }\label{fig.shor5}        
\end{figure}

\begin{figure}
\epsfig{file=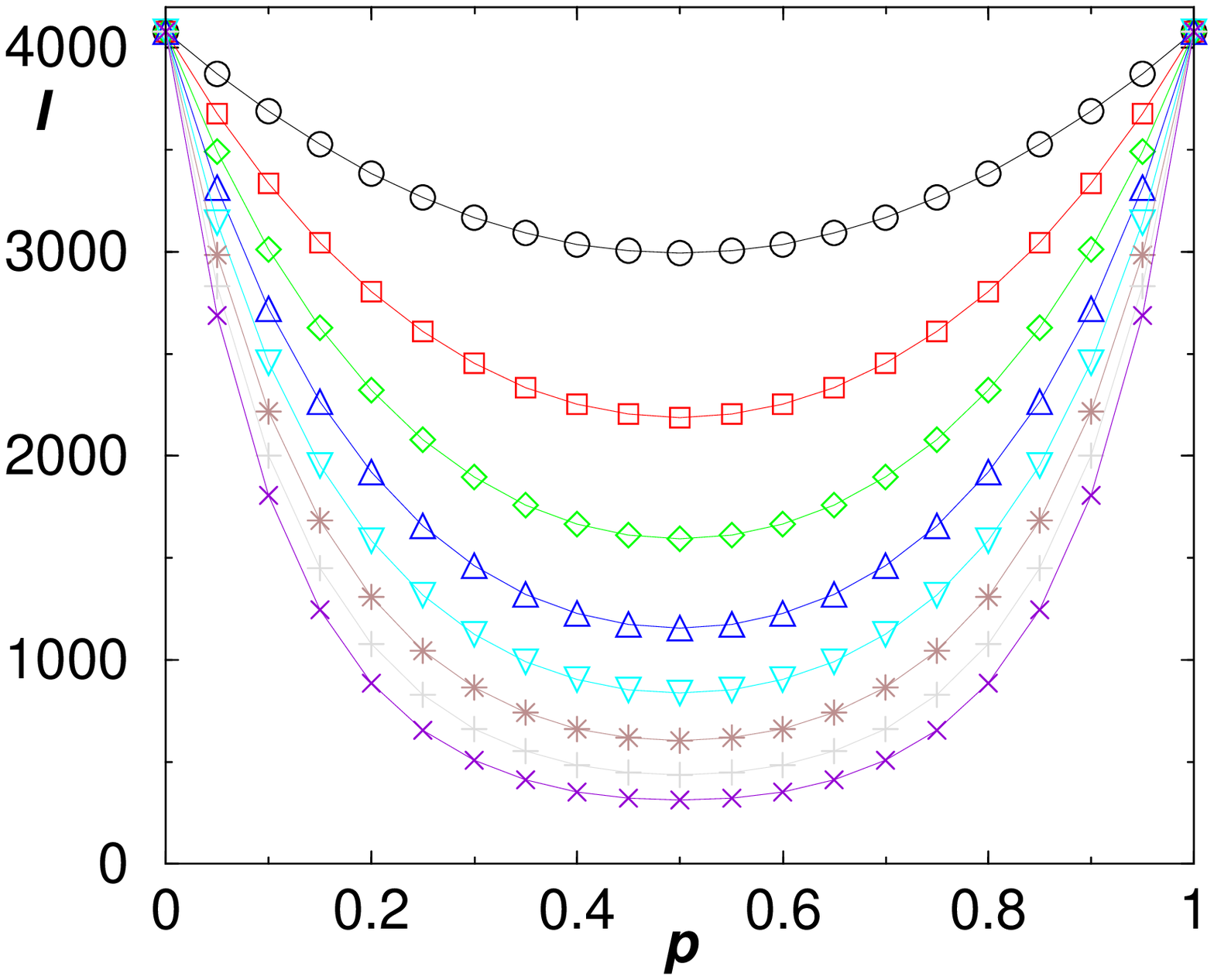,width=5cm,angle=0}
\epsfig{file=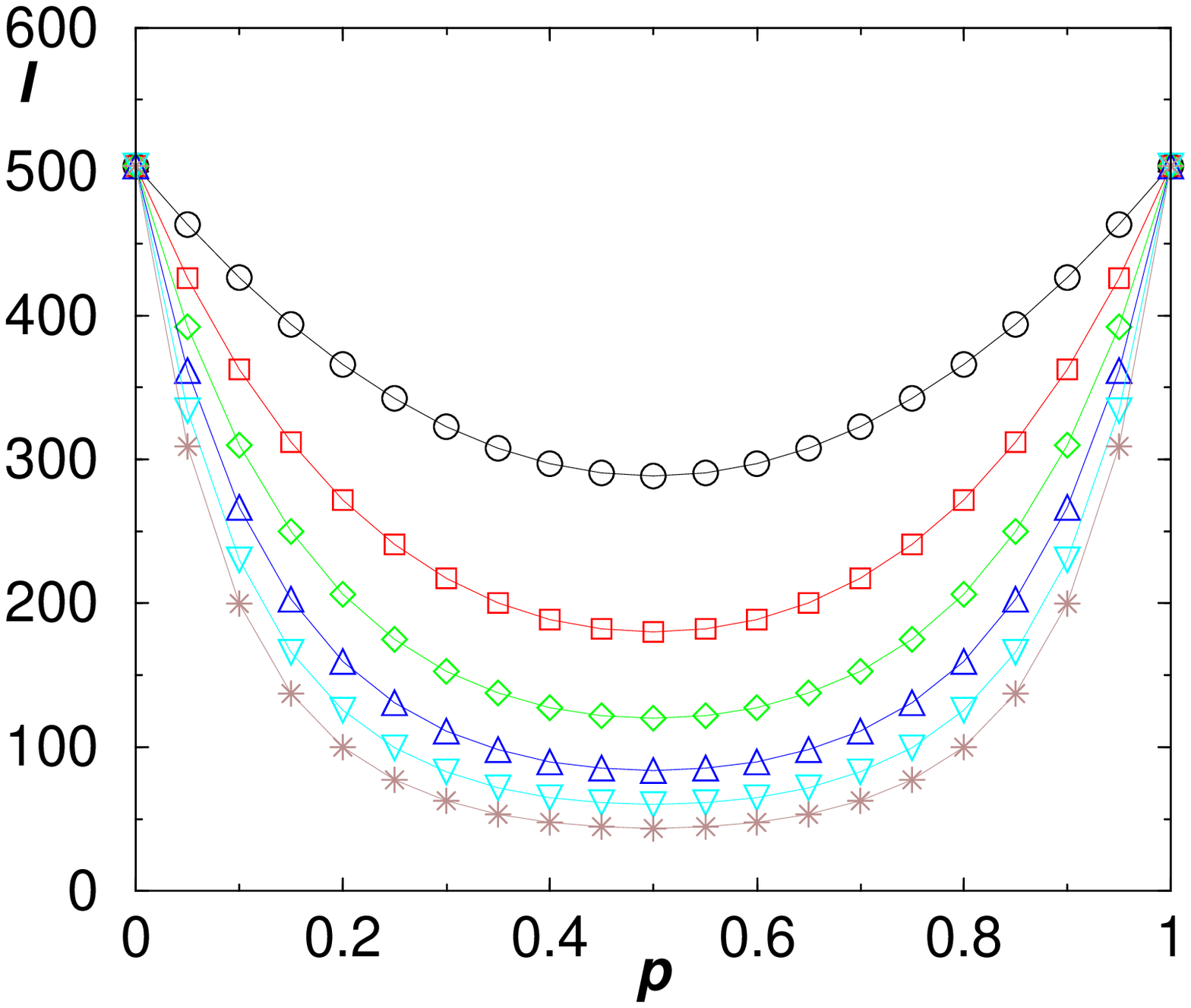,width=5cm,angle=0}
\epsfig{file=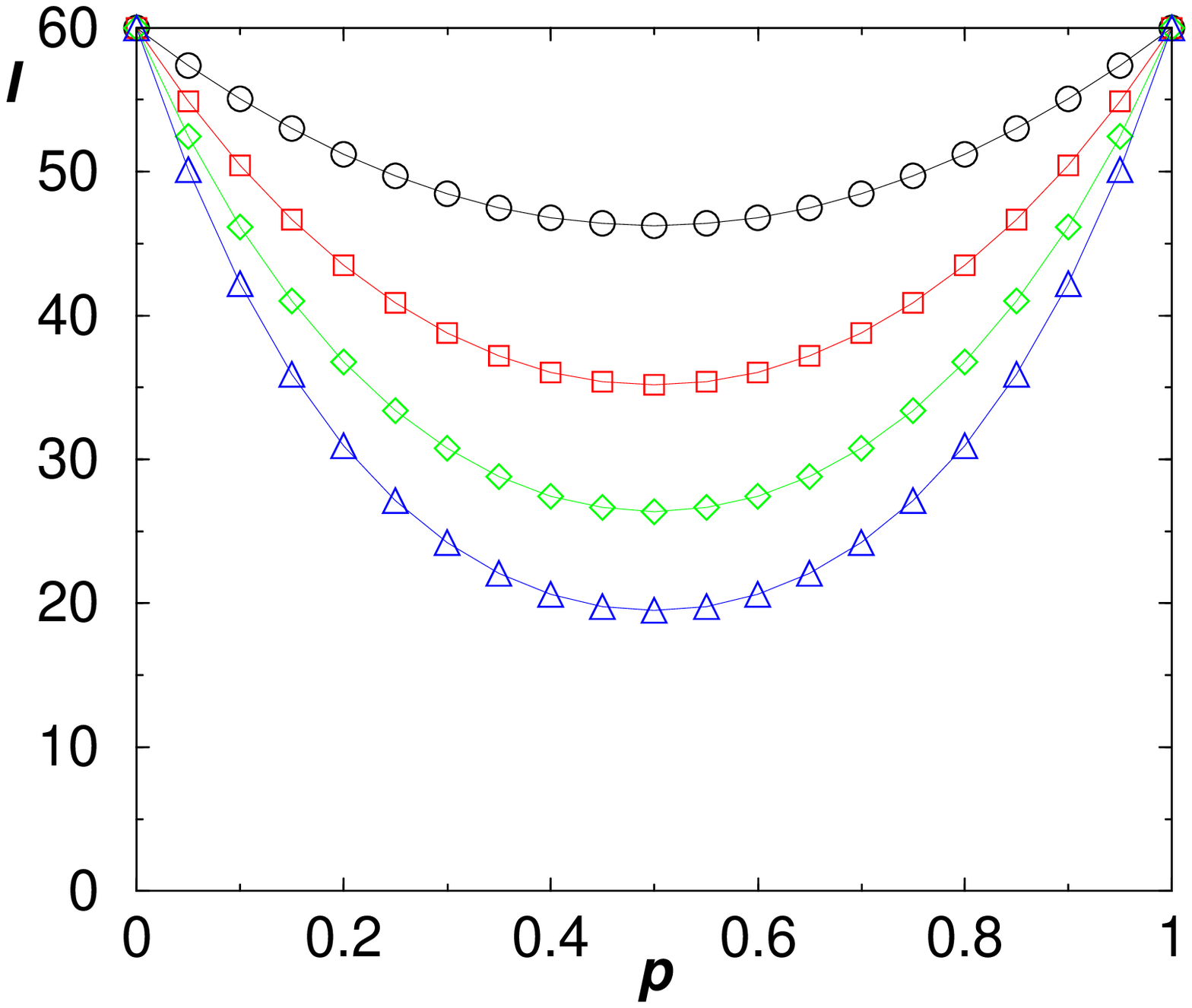,width=5cm,angle=0}
\caption{(Color online) Actually used interference in the Shor
  algorithm with decoherence through bit--flips during the first
  Walsh-Hadamard transformation, as function of the bit--flip probability
  $p$ for 
$n=12$ (left), $n=9$ (center), $n=6$ (right). The corresponding success
  probability $S$ 
is constant equal to $S=1$ for all values of $p$ (data not shown); symbols
  as in Fig.~\ref{fig.shor5}. 
  }\label{fig.shor6}        
\end{figure}

Figures \ref{fig.shor5}-\ref{fig.shor6} show the result of decoherence due
to bit-flips in the Shor algorithm for $n=12$, $n=9$ and $n=6$ qubits.
As for the Grover algorithm, bit-flips are performed after each of 
the initial Hadamard gates.  However, as these Hadamard gates concern
only one of the registers, the decoherence process affects only
the first two-thirds of qubits.   
The curves show the effect of decoherence on a growing number of qubits,
from $n_f=1$ to $n_f=2/3n$, with data averaged over the choice of 
the $n_f$ affected qubits.  
The success probability $S$ is computed as in eq.(\ref{sucprob}) where now
$|\psi_{i}^{err}|^2$ is replaced by the probability for state $i$ in the
final mixed state. 

The success probability is constant equal to $1$ for all values
of $p$ (data not shown). In contrast, both potentially available and 
actually used interference are strongly affected by the decoherence.
Both quantities decrease from their maximum value at $p=0$ and $p=1$
to a minimum at $p=0.5$, the potentially available interference
decreasing faster.  However, the interference never goes down to zero in this
setting, as can be seen in the insets of Figs.\ref{fig.shor5}\ref{fig.shor6},
contrary to the case of the Grover algorithm above.  

\begin{figure}
\epsfig{file=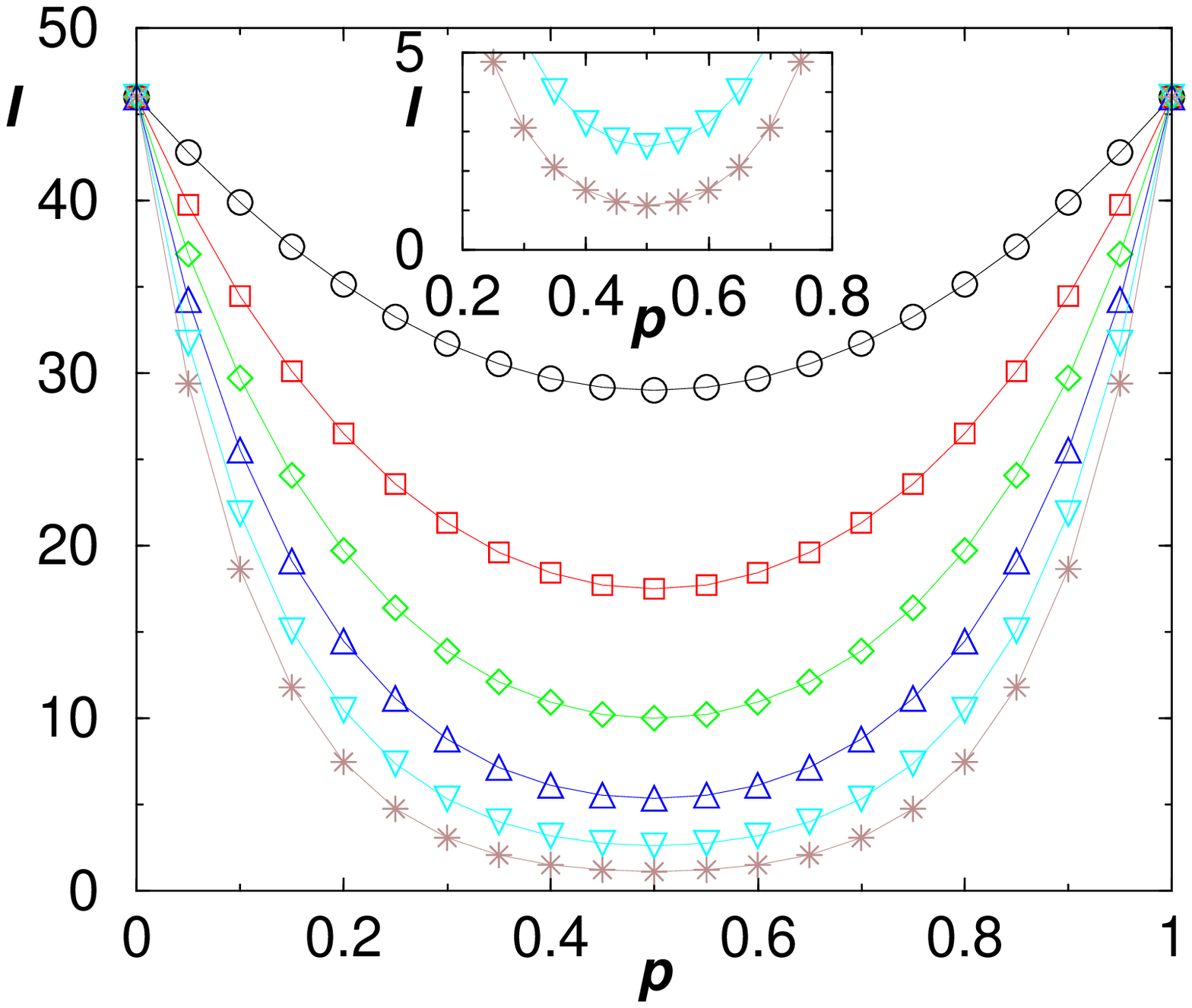,width=5cm,angle=0}
\epsfig{file=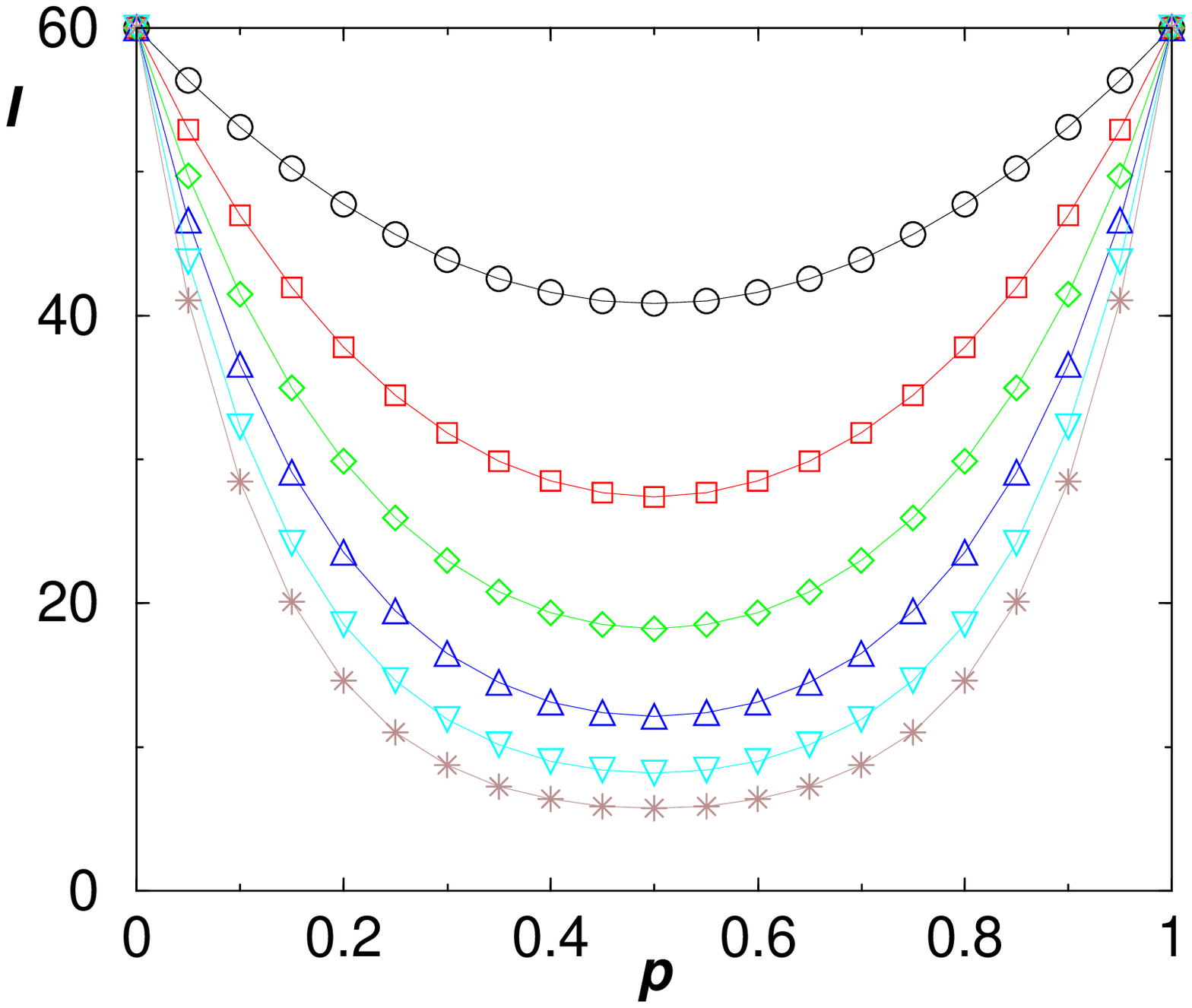,width=5cm,angle=0}
\epsfig{file=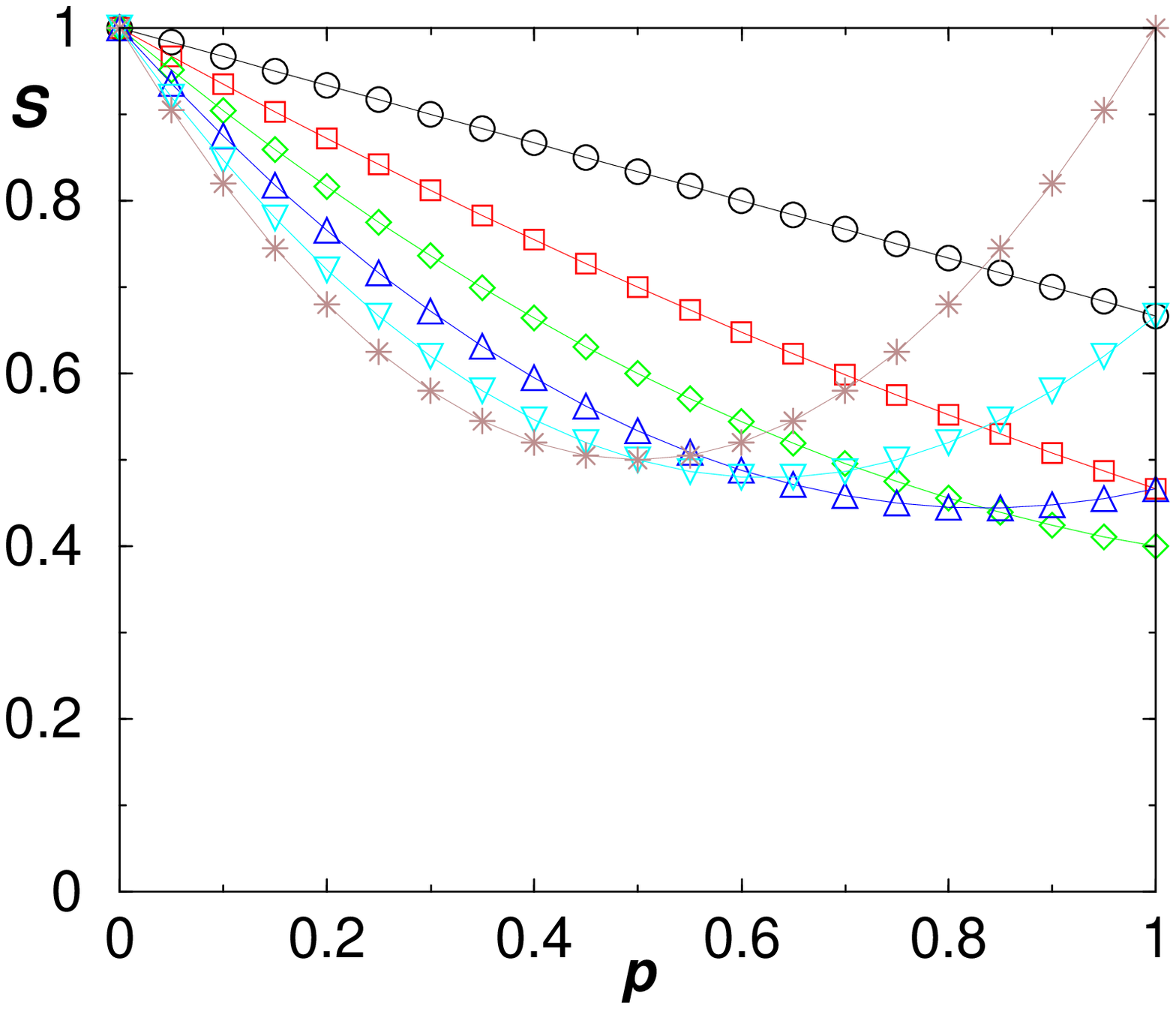,width=5cm,angle=0}
\caption{(Color online) Interference and success probability in the Shor
  algorithm with decoherence through bit--flips during the first
  Walsh-Hadamard transformation, as function of the bit--flip probability
  $p$ after each Hadamard gate, with all $n$ qubits flipped, for
$n=6$: 
potentially available interference (left), 
actually used interference (center), success probability (right).
Inset on the left is a close-up close to the minimum;  symbols
  as in Fig.~\ref{fig.shor5}.
  }\label{fig.shor7}        
\end{figure}

In the Figs.\ref{fig.shor5}-\ref{fig.shor6}, only two-thirds of
the qubits are affected by the decoherence.  This may appear to be the main
reason why the potentially available interference does not decrease to zero
in presence of bit-flip  
decoherence, contrary to the case above with the Grover algorithm. 
In order to investigate in more details this question,
we studied the interference produced when {\em all} $n$ qubits are affected
by bit-flip decoherence.  The results are shown in Fig.~\ref{fig.shor7}
for $n=6$.
Although the two types of interference decrease faster than in 
Figs.\ref{fig.shor5}-\ref{fig.shor6}, none of them reaches 
exactly zero over the whole interval of $p$ values.  Contrary to the case
of Figs.\ref{fig.shor5}-\ref{fig.shor6}, 
the success probability is now strongly affected 
by the decoherence and is not preserved: the initial superposed state
is not any more protected against this type of decoherence, since the
second register is not supposed to be in a equal superposition state
in the exact algorithm.
The data displayed in Figs. \ref{fig.shor5}-\ref{fig.shor7}
suggest that for Shor's algorithm, although it is possible
to decrease the interference by a large amount while keeping the success
probability 
constant, it does not seem possible to perform efficiently the computation
with zero interference.

\begin{figure}
\epsfig{file=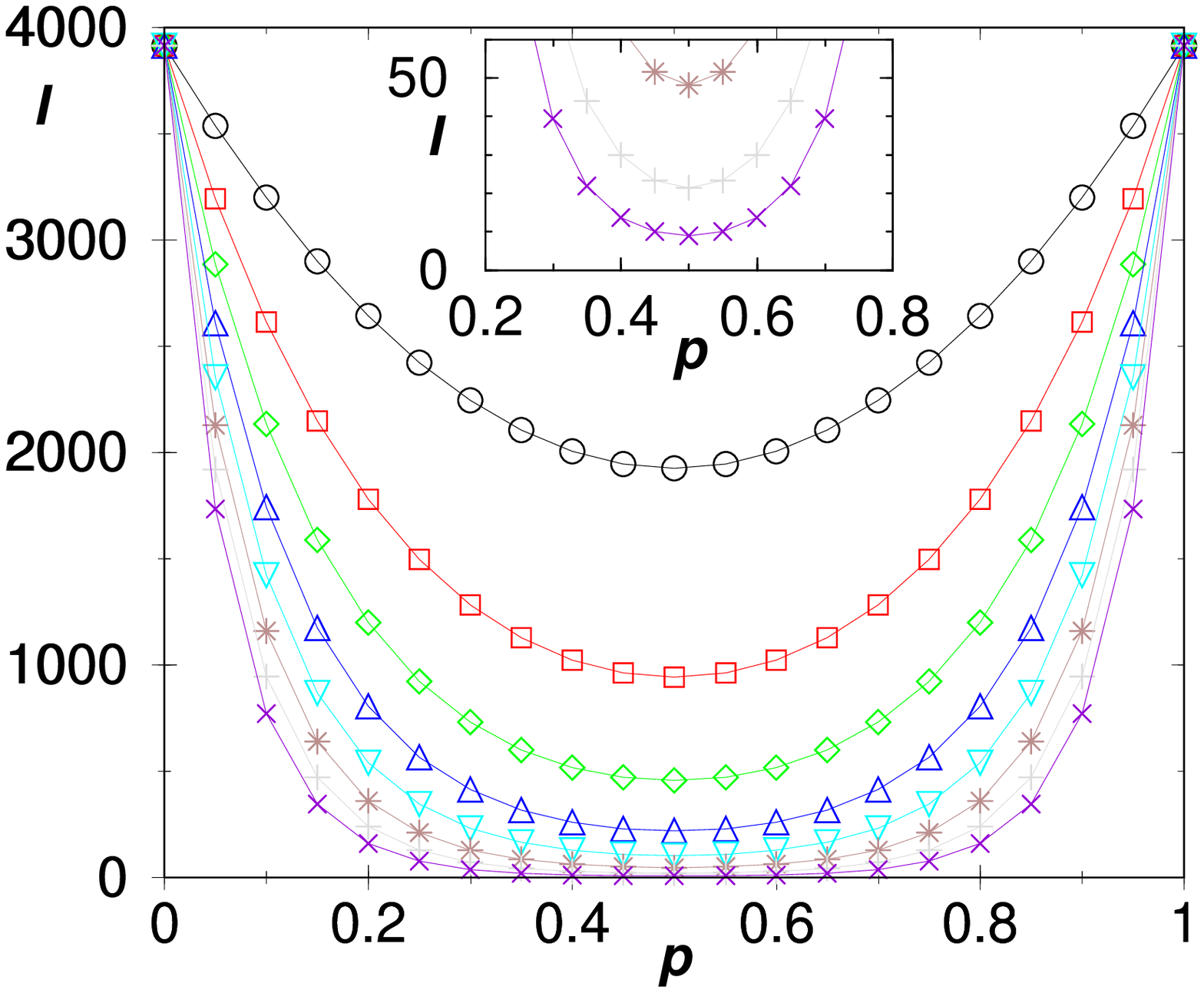,width=5cm,angle=0}
\epsfig{file=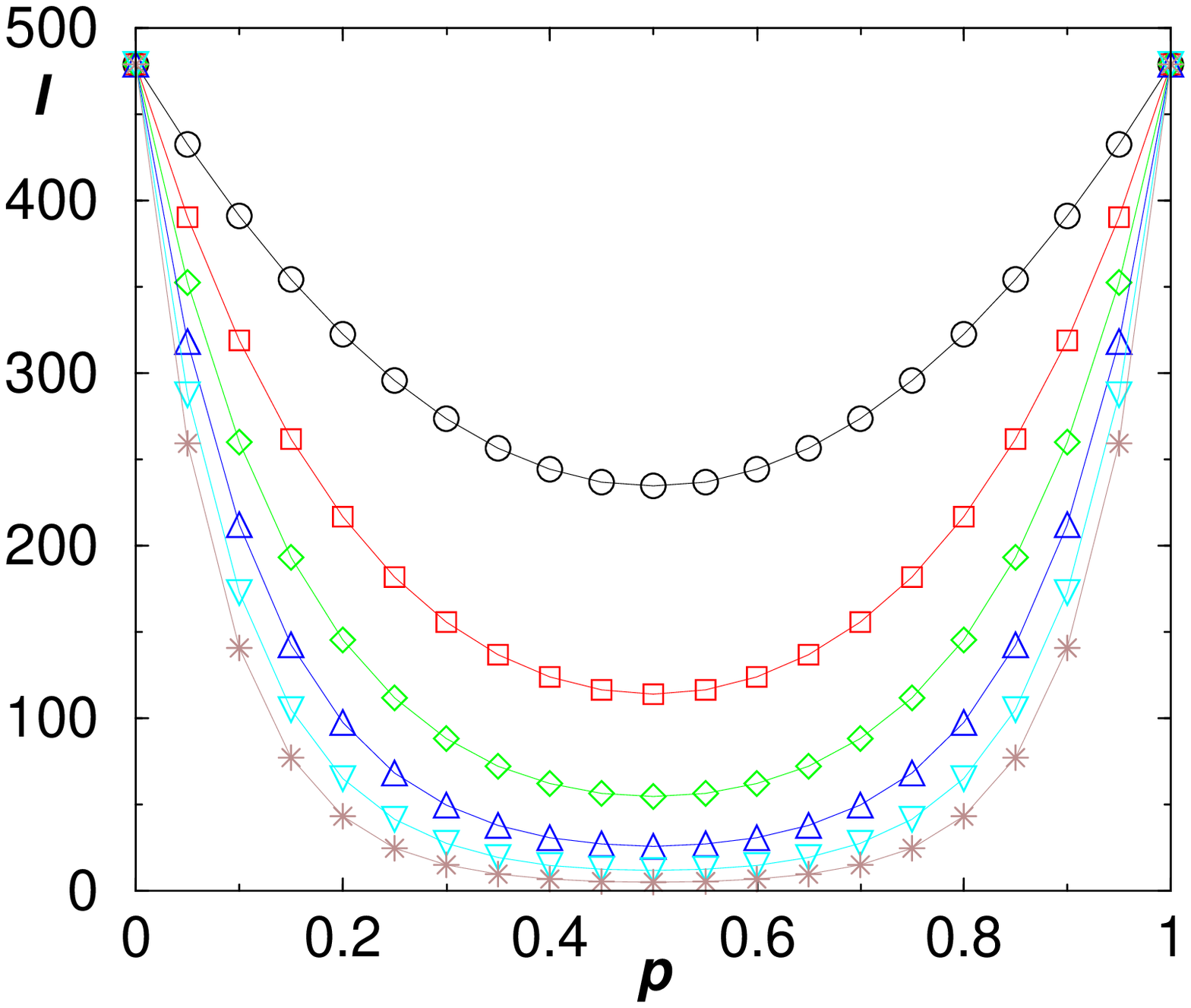,width=5cm,angle=0}
\epsfig{file=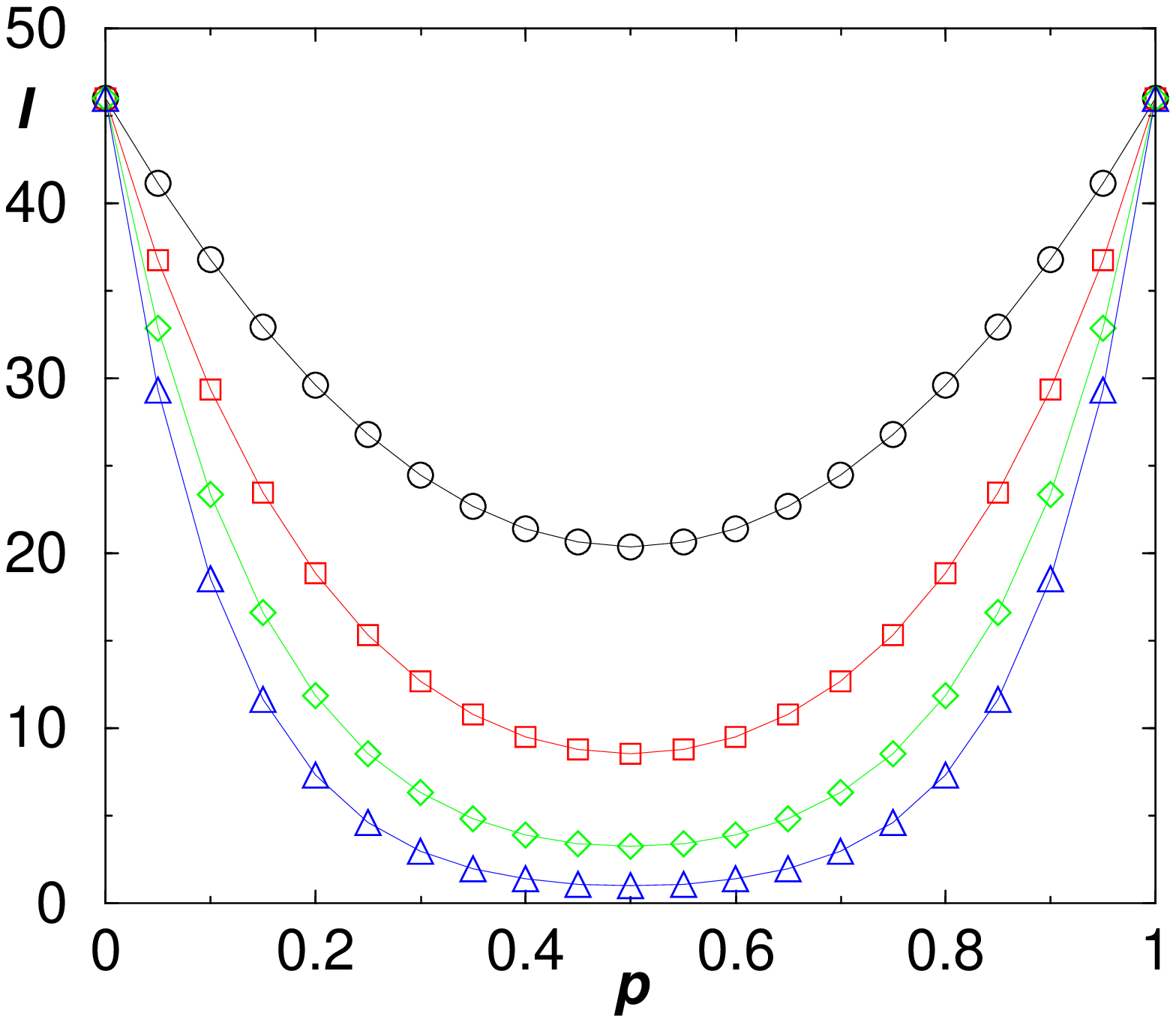,width=5cm,angle=0}
\caption{(Color online) Potentially available interference in the Shor
  algorithm with decoherence through phase--flips during the first
  Walsh-Hadamard transformation, as function of the phase--flip probability
  $p$ after each Hadamard gate, for 
$n=12$ (left), $n=9$ (center), $n=6$ (right). 
Inset on the left is a close-up of the 
case $n=12$ close to the minimum; symbols
  as in Fig.~\ref{fig.shor5}.
  }\label{fig.shor8}        
\end{figure}

\begin{figure}
\epsfig{file=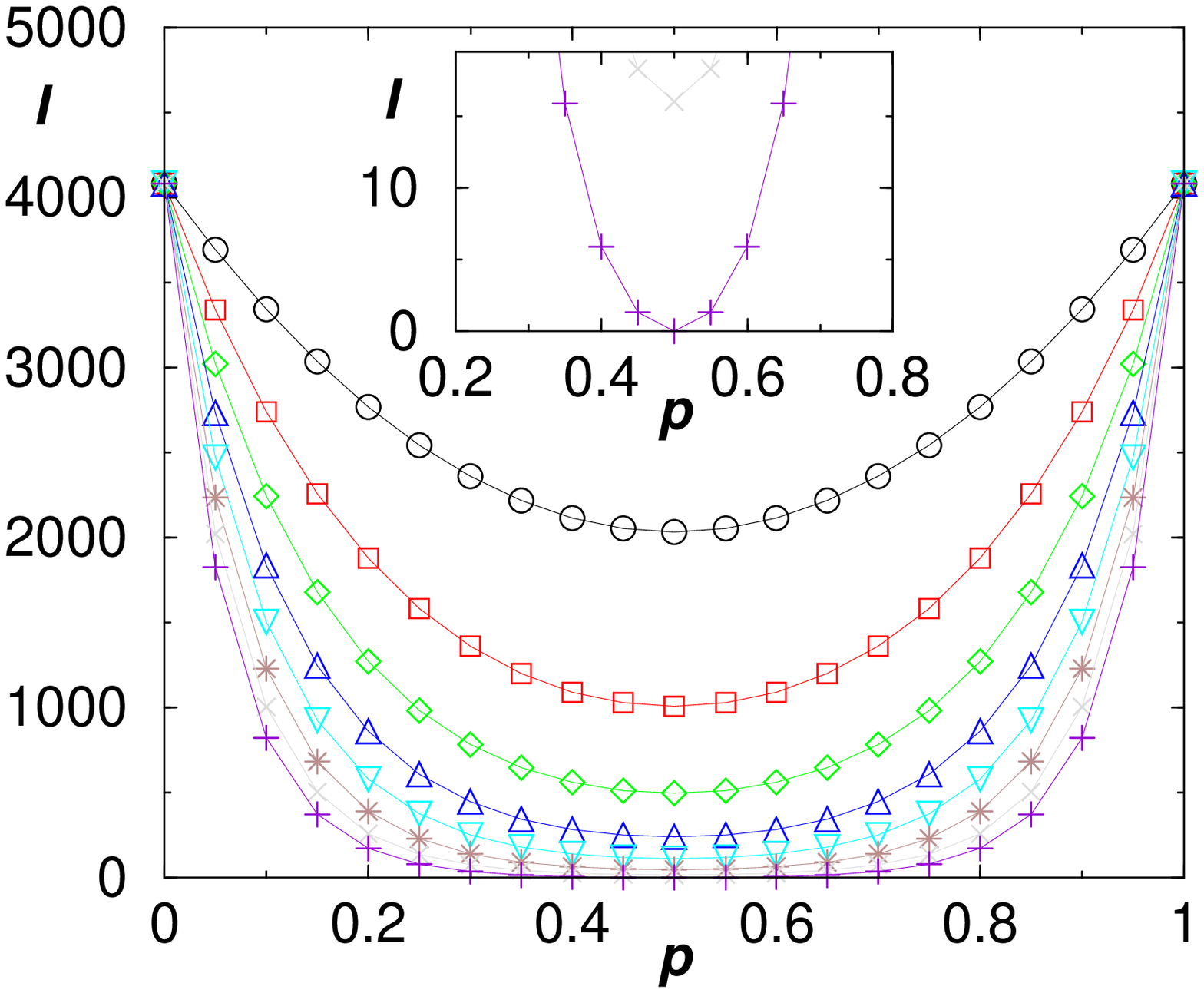,width=5cm,angle=0}
\epsfig{file=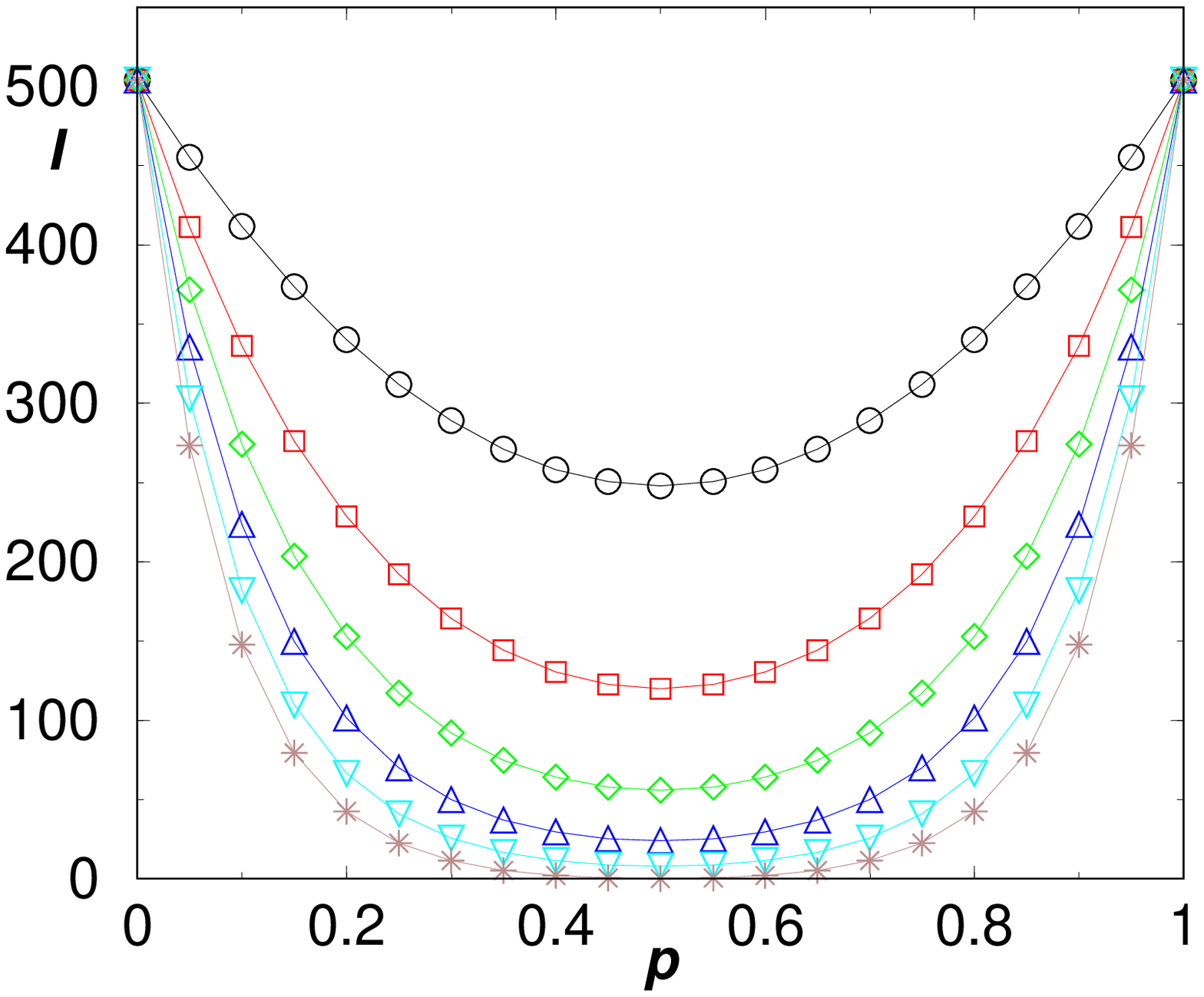,width=5cm,angle=0}
\epsfig{file=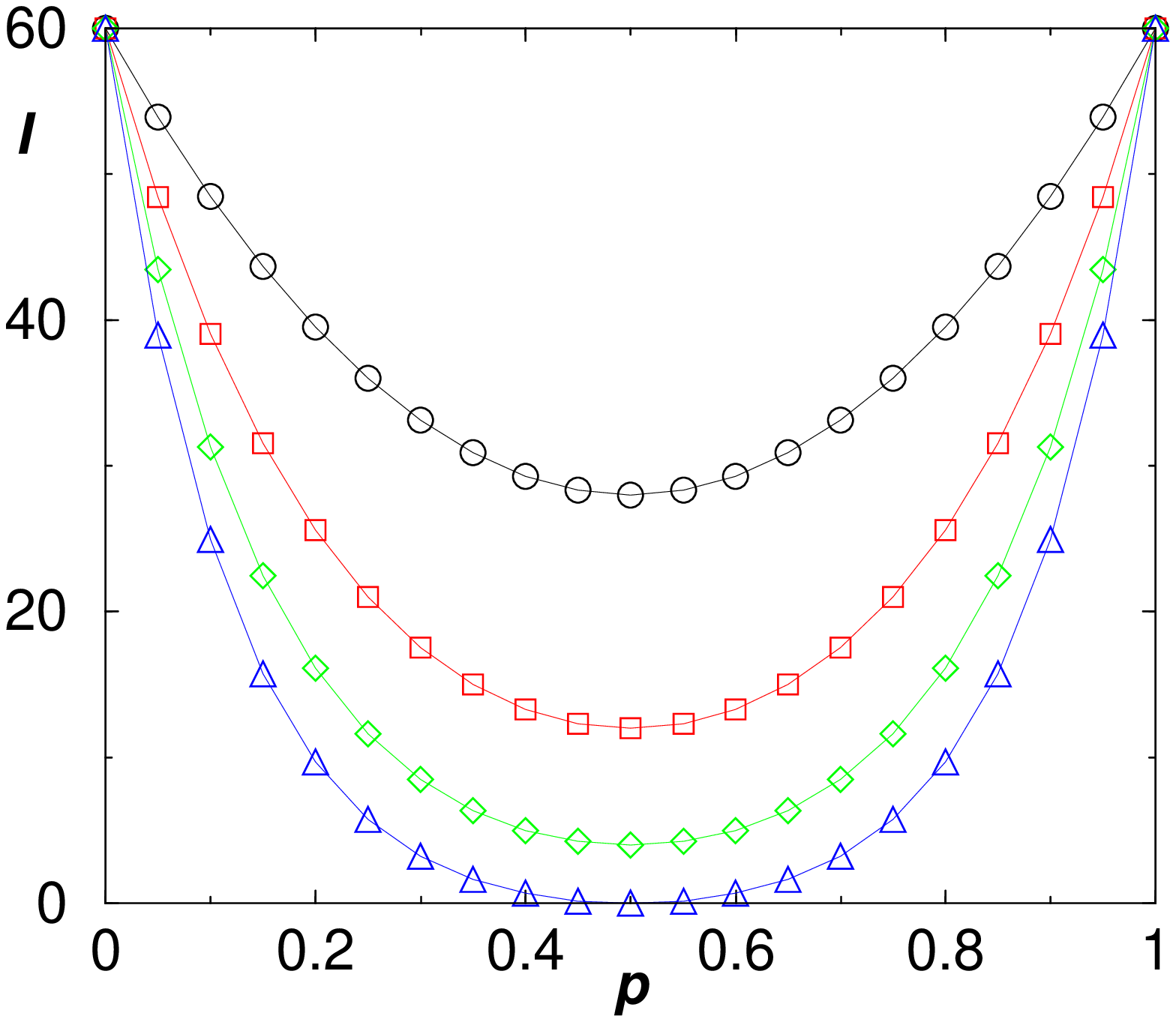,width=5cm,angle=0}
\caption{(Color online) Actually used interference in the Shor
  algorithm with decoherence through phase--flips during the first
  Walsh-Hadamard transformation, as function of the phase--flip probability
  $p$ after each Hadamard gate for 
$n=12$ (left), $n=9$ (center), $n=6$ (right).
Inset on the left is a close-up of the 
case $n=12$ close to the minimum, showing that the value $\cI=0$
is indeed reached for $p=0.5$.
  }\label{fig.shor9}        
\end{figure}

\begin{figure}
\epsfig{file=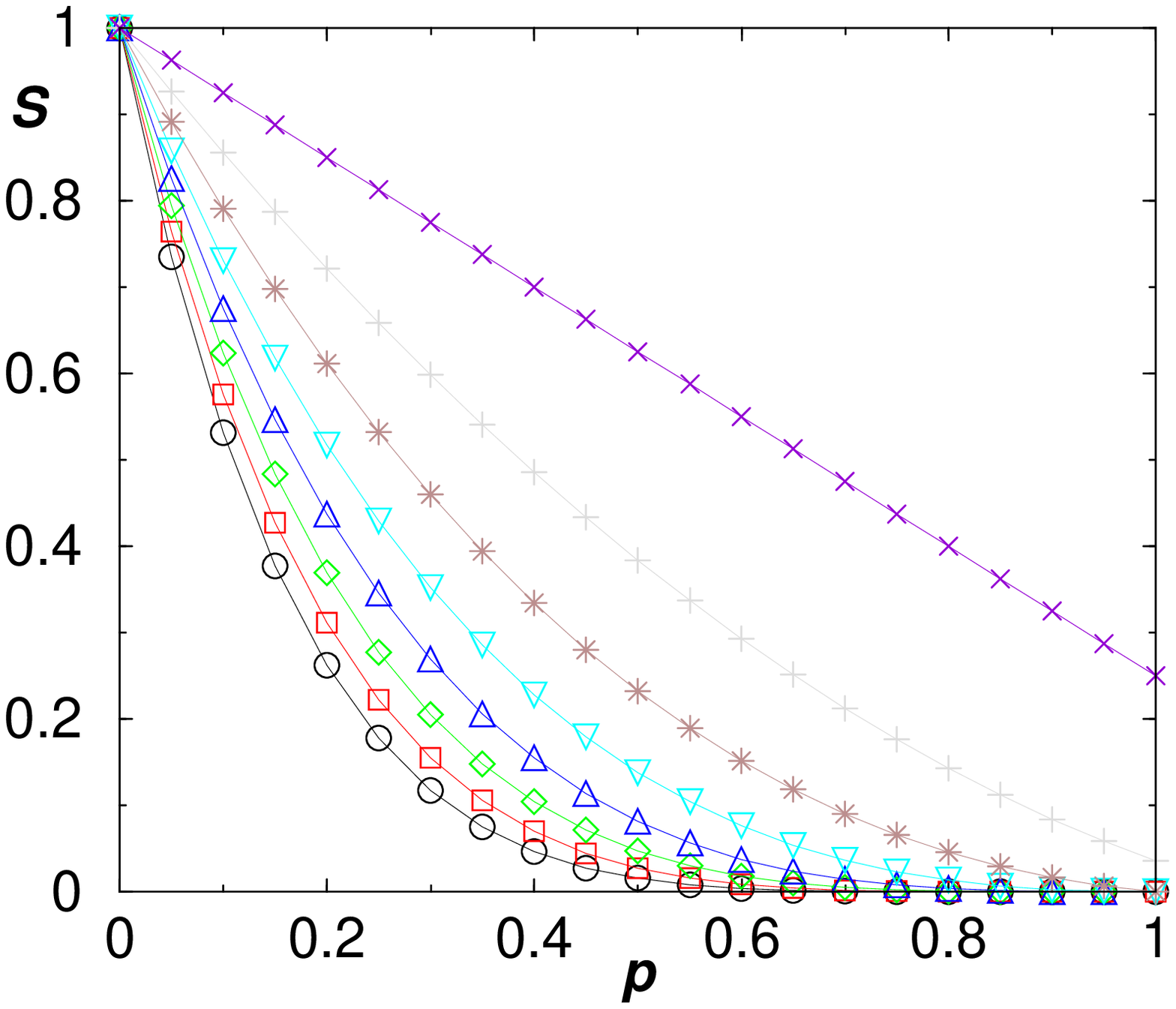,width=4cm,angle=0}
\epsfig{file=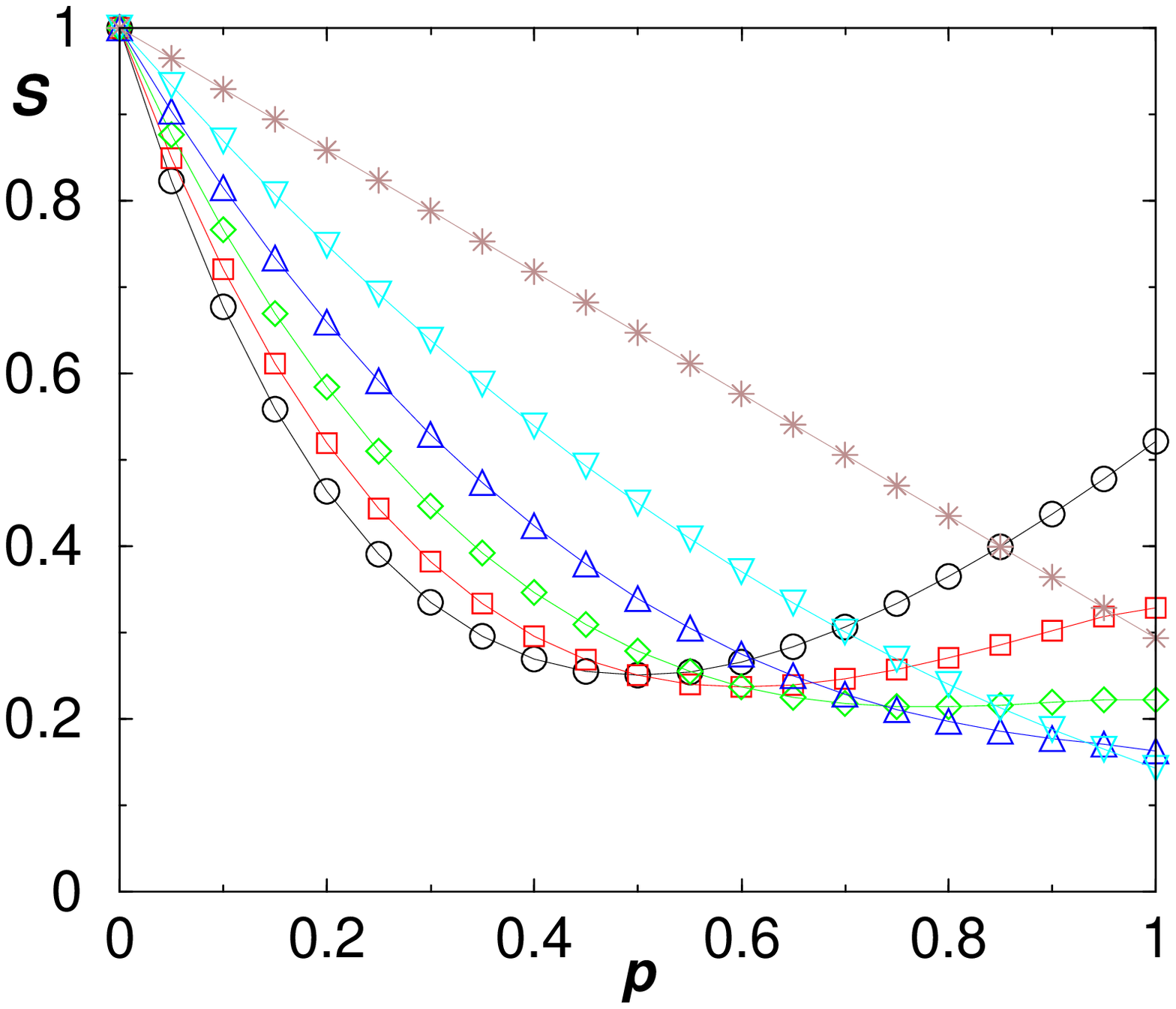,width=4cm,angle=0}
\epsfig{file=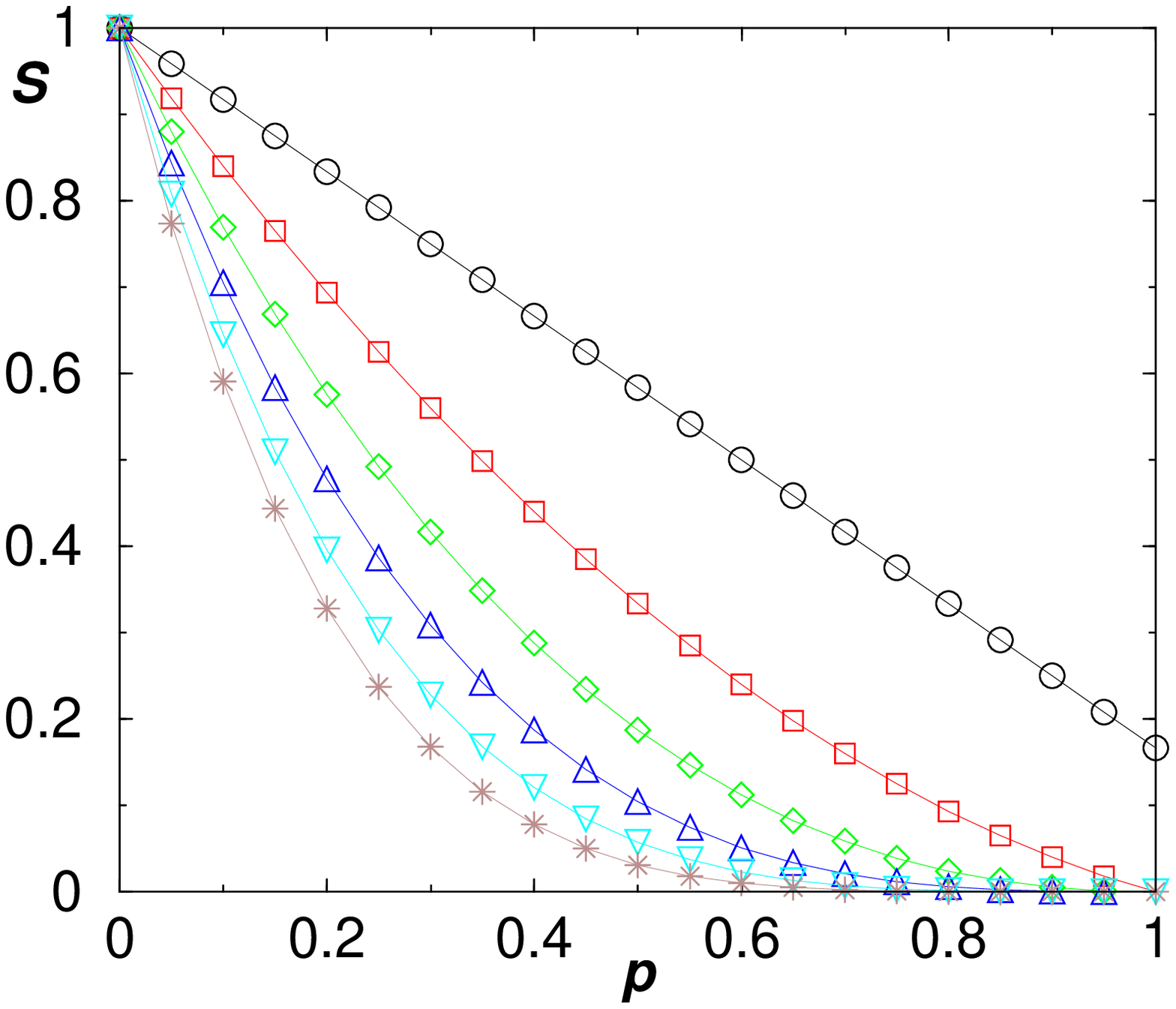,width=4cm,angle=0}
\epsfig{file=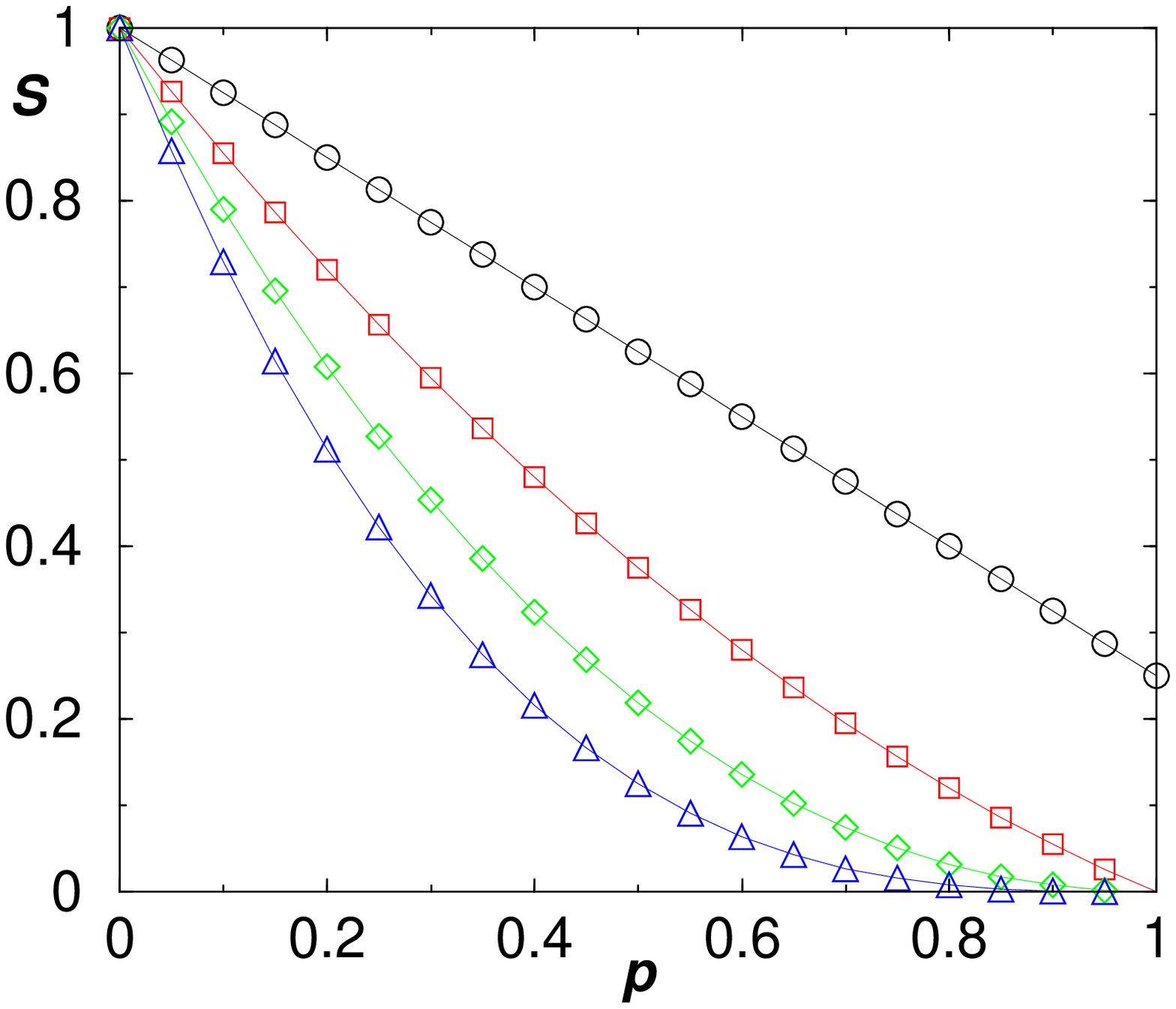,width=4cm,angle=0}
\caption{(Color online) Success Probability in the Shor
  algorithm with decoherence through phase--flips during the first
  Walsh-Hadamard transformation, as function of the phase--flip probability
  $p$ after each Hadamard gate for 
$n=12$ (left), $n=9$ $a=3$ (center left), 
$n=9$ $a=6$ (center right), $n=6$ (right); symbols
  as in Fig.~\ref{fig.shor5}.
  }\label{fig.shor10}        
\end{figure}

Figures \ref{fig.shor8}-\ref{fig.shor10} display data obtained
for decoherence through phase-flips. As before, phase flips are introduced
with probability $p$ on qubits of the first register after application
of the Hadamard gates.  The data displayed in 
Figs.\ref{fig.shor8}-\ref{fig.shor9}
show that interference, both 
potentially available and actually used, decreases to a minimum at $p=0.5$.
The minimum is lower than in the case of bit-flips, and reaches zero 
for actually used interference.  In the case of potentially available 
interference, some residual interference is still present when all qubits
of the first register are affected. Figure \ref{fig.shor10} shows that in
contrast to the case of bit-flip errors, success probability is strongly
affected for  
phase-flip errors.  It decreases with the number of qubits affected and
the value of $p$ until the algorithm is totally destroyed. Comparison 
with the Figs.\ref{fig.shor8}-\ref{fig.shor9} 
shows that the increase of interference
between $p=0.5$ and $p=1$ is not reflected in a similar increase
in success probability.  the interference produced in this case is useless
and does 
not serve the algorithm.  It is similar to the one found for random algorithms 
in \cite{Arnaud07}, where it was shown that random algorithms produce
on average an interference close to the maximum value.  The case
$n=9$, $a=3$ is peculiar: in this particular instance (second figure in 
Fig.~\ref{fig.shor10}), after an initial decrease
for small values of $p$, the success probability increases for larger
values of $p$, 
although it never reaches values close to one.  We think this is due to
the fact that in this case the period is not a power of two, and therefore
the final wavefunction is composed of broad peaks which for such small
sizes have a significant projection on many basis states of the Hilbert
space.  A random-type wavefunction produced by the destroyed Shor algorithm 
therefore has a much larger projection on such state than on a state composed
of sharp $\delta$ peaks as in the two other values of $n$.  To sustain this
hypothesis, we computed the success probability for $n=9$ and $a=6$, where
the period 
does divide the Hilbert space dimension, and the final wave function is
composed of $\delta$-peaks.  In this case (last figure of 
Fig.~\ref{fig.shor10}) the success probability indeed goes to zero for large
$p$  values. 

\section{Conclusions} 
In this paper we have investigated the success probability of quantum
algorithms in relation with the interference produced by the algorithms. To
this end we subjected Grover's search algorithm and Shor's factoring
algorithm to different kinds of errors, namely systematic unitary errors,
random unitary errors, and decoherence due to bit flips or phase flips.  
The study of systematic unitary errors showed that in both algorithms the
controlled 
destruction of interference goes hand in hand with the decay of the
success probability. This reinforced the idea that interference is an
important ingredient necessary for the functioning of these algorithms. 

The
case of random unitary errors shows, however, that a large amount of
interference is by no means sufficient for the success of a quantum
algorithm, since in some cases the interference increases with decreasing
success probability. This effect is  particularly pronounced for the
actually used interference in the case of Grover's algorithm, where the
interference increases from about two i-bits for the unperturbed algorithm
to an amount of the order $n$ i-bits, close to the maximum possible value,
for sufficiently strong errors. The success probability may decrease even
below the classical value corresponding to unbiased random guessing,
meaning that the interference has become destructive. This can also
be understood in the context of the recent result that a randomly chosen 
quantum algorithm leads with very high probability to an
amount of interference close to the maximum value \cite{Arnaud07}, such that
randomizing a given quantum algorithm with limited interference is very
likely to increase the interference, even if the algorithm itself is
destroyed in the process. Thus, as to be expected, interference needs to be
exploited in the proper way to be useful.  

The study of decoherence led to more complex results. 
Phase-flip errors destroy interference and in parallel decrease the success
probability, in the same way as systematic errors.  Interference decreases
for small error probabilities, and reappears again as destructive 
  interference for error probabilities approaching one, leading to
  success probabilities which can even drop below the classical
  value. 
In contrast, bit-flip
errors performed after each initial Hadamard gate do not reduce the success 
probability of the algorithm, while affecting the interference
produced.  In the case of Grover's algorithm, the potentially 
available interference can even go all the way down to zero while
the performance of the algorithm is unaffected.  This surprising result
is due to the symmetry of the equipartitioned state used as initial state
in the Grover's algorithm, which is invariant under bit-flips.
It should be remarked, however, that the actually used interference does
not go to zero for Grover's algorithm. As concerns Shor's algorithm, the
bit flips destroy part but not all of the interference, 
both potentially available and actually used,
while also keeping constant the success probability of the algorithm.
These results show that in general it is possible to reduce substantially
the interference produced while keeping the efficiency of the algorithm. 
Grover's algorithm seems to run correctly with zero potentially available
interference, but not with zero actually used interference.  In contrast,
in none of our simulation Shor's algorithm was found to run efficiently
without some interference left, potentially available or actually used.

\begin{acknowledgments}
      We thank IDRIS in Orsay and CALMIP in Toulouse 
for use of their computers.
This work was supported in part by the Agence  
National de la Recherche (ANR), project INFOSYSQQ, and the EC IST-FET
project EUROSQIP. 
\end{acknowledgments}

\bibliography{../mybibs_bt}

\begin{thebibliography}{20}
\expandafter\ifx\csname natexlab\endcsname\relax\def\natexlab#1{#1}\fi
\expandafter\ifx\csname bibnamefont\endcsname\relax
  \def\bibnamefont#1{#1}\fi
\expandafter\ifx\csname bibfnamefont\endcsname\relax
  \def\bibfnamefont#1{#1}\fi
\expandafter\ifx\csname citenamefont\endcsname\relax
  \def\citenamefont#1{#1}\fi
\expandafter\ifx\csname url\endcsname\relax
  \def\url#1{\texttt{#1}}\fi
\expandafter\ifx\csname urlprefix\endcsname\relax\def\urlprefix{URL }\fi
\providecommand{\bibinfo}[2]{#2}
\providecommand{\eprint}[2][]{\url{#2}}

\bibitem[{\citenamefont{Cleve et~al.}(1998)\citenamefont{Cleve, Ekert,
  Macchiavello, and Mosca}}]{Cleve98}
\bibinfo{author}{\bibfnamefont{R.}~\bibnamefont{Cleve}},
  \bibinfo{author}{\bibfnamefont{A.}~\bibnamefont{Ekert}},
  \bibinfo{author}{\bibfnamefont{C.}~\bibnamefont{Macchiavello}},
  \bibnamefont{and} \bibinfo{author}{\bibfnamefont{M.}~\bibnamefont{Mosca}},
  \bibinfo{journal}{Proc.Royal Soc. A} \textbf{\bibinfo{volume}{454}},
  \bibinfo{pages}{339} (\bibinfo{year}{1998}).

\bibitem[{\citenamefont{Bennett and DiVincenzo}(2000)}]{Bennett00}
\bibinfo{author}{\bibfnamefont{C.~H.} \bibnamefont{Bennett}} \bibnamefont{and}
  \bibinfo{author}{\bibfnamefont{D.~P.} \bibnamefont{DiVincenzo}},
  \bibinfo{journal}{Nature} \textbf{\bibinfo{volume}{404}},
  \bibinfo{pages}{247} (\bibinfo{year}{2000}).

\bibitem[{\citenamefont{Beaudry et~al.}(2005)\citenamefont{Beaudry, Fernandez,
  and Holzer}}]{Beaudry05}
\bibinfo{author}{\bibfnamefont{M.}~\bibnamefont{Beaudry}},
  \bibinfo{author}{\bibfnamefont{J.~M.} \bibnamefont{Fernandez}},
  \bibnamefont{and} \bibinfo{author}{\bibfnamefont{M.}~\bibnamefont{Holzer}},
  \bibinfo{journal}{Theor. Comp. Science} \textbf{\bibinfo{volume}{345}},
  \bibinfo{pages}{206} (\bibinfo{year}{2005}).

\bibitem[{\citenamefont{Shor}(1994)}]{Shor94}
\bibinfo{author}{\bibfnamefont{P.~W.} \bibnamefont{Shor}}, in
  \emph{\bibinfo{booktitle}{Proc. 35th Annu. Symp. Foundations of Computer
  Science (ed. Goldwasser, S.), {\em p.~124-134}}} (\bibinfo{publisher}{IEEE
  Computer Society, Los Alamitos, CA}, \bibinfo{year}{1994}).

\bibitem[{\citenamefont{Grover}(1997)}]{Grover97}
\bibinfo{author}{\bibfnamefont{L.~K.} \bibnamefont{Grover}},
  \bibinfo{journal}{Phys. Rev. Lett.} \textbf{\bibinfo{volume}{79}},
  \bibinfo{pages}{325} (\bibinfo{year}{1997}).

\bibitem[{\citenamefont{Boyer et~al.}(1998)\citenamefont{Boyer, Brassard,
  H{\o}yer, and Tapp}}]{Boyer98}
\bibinfo{author}{\bibfnamefont{M.}~\bibnamefont{Boyer}},
  \bibinfo{author}{\bibfnamefont{G.}~\bibnamefont{Brassard}},
  \bibinfo{author}{\bibfnamefont{P.}~\bibnamefont{H{\o}yer}}, \bibnamefont{and}
  \bibinfo{author}{\bibfnamefont{A.}~\bibnamefont{Tapp}},
  \bibinfo{journal}{Fortschr. Phys.} \textbf{\bibinfo{volume}{46}},
  \bibinfo{pages}{493} (\bibinfo{year}{1998}).

\bibitem[{\citenamefont{Brassard et~al.}(2002)\citenamefont{Brassard,
  P.~H{\o}yer, and Tapp}}]{Brassard02}
\bibinfo{author}{\bibfnamefont{G.}~\bibnamefont{Brassard}},
  \bibinfo{author}{\bibfnamefont{M.~M.} \bibnamefont{P.~H{\o}yer}},
  \bibnamefont{and} \bibinfo{author}{\bibfnamefont{A.}~\bibnamefont{Tapp}}, in
  \emph{\bibinfo{booktitle}{Quantum Computation and Quantum Information: A
  Millenium Volume}} (\bibinfo{publisher}{edited by S. J. Lomonaco, Jr. and H.
  E. Brandt}, \bibinfo{year}{2002}), (AMS, Contemporary Mathematics Series Vol.
  305).

\bibitem[{\citenamefont{B.Georgeot}(2004)}]{Georgeot04}
\bibinfo{author}{\bibnamefont{B.Georgeot}}, \bibinfo{journal}{Phys. Rev. A}
  \textbf{\bibinfo{volume}{69}}, \bibinfo{pages}{032301}
  (\bibinfo{year}{2004}).

\bibitem[{\citenamefont{Deutsch}(1985)}]{Deutsch85}
\bibinfo{author}{\bibfnamefont{D.}~\bibnamefont{Deutsch}},
  \bibinfo{journal}{Proc. Roy. Soc. Lond. A} \textbf{\bibinfo{volume}{400}},
  \bibinfo{pages}{97} (\bibinfo{year}{1985}).

\bibitem[{\citenamefont{Jozsa and Linden}(2003)}]{Jozsa03}
\bibinfo{author}{\bibfnamefont{R.}~\bibnamefont{Jozsa}} \bibnamefont{and}
  \bibinfo{author}{\bibfnamefont{N.}~\bibnamefont{Linden}},
  \bibinfo{journal}{Proc. R. Soc. Lond. A} \textbf{\bibinfo{volume}{459}},
  \bibinfo{pages}{2011} (\bibinfo{year}{2003}).

\bibitem[{\citenamefont{Bruss}(2002)}]{Bruss01}
\bibinfo{author}{\bibfnamefont{D.}~\bibnamefont{Bruss}}, \bibinfo{journal}{J.
  Math. Phys.} \textbf{\bibinfo{volume}{43}}, \bibinfo{pages}{4237}
  (\bibinfo{year}{2002}).

\bibitem[{\citenamefont{De et~al.}()\citenamefont{De, Sen, Lewenstein, and
  Sanpera}}]{DeSLS05}
\bibinfo{author}{\bibfnamefont{A.}~\bibnamefont{De}},
  \bibinfo{author}{\bibfnamefont{U.}~\bibnamefont{Sen}},
  \bibinfo{author}{\bibfnamefont{M.}~\bibnamefont{Lewenstein}},
  \bibnamefont{and} \bibinfo{author}{\bibfnamefont{A.}~\bibnamefont{Sanpera}},
  \eprint{quant-ph/0508032}.

\bibitem[{\citenamefont{Braun and Georgeot}(2006)}]{Braun06}
\bibinfo{author}{\bibfnamefont{D.}~\bibnamefont{Braun}} \bibnamefont{and}
  \bibinfo{author}{\bibfnamefont{B.}~\bibnamefont{Georgeot}},
  \bibinfo{journal}{Phys. Rev. A} \textbf{\bibinfo{volume}{73}},
  \bibinfo{pages}{022314} (\bibinfo{year}{2006}).

\bibitem[{\citenamefont{Aharonov et~al.}()\citenamefont{Aharonov, Landau, and
  Makowsky}}]{Aharonov06}
\bibinfo{author}{\bibfnamefont{D.}~\bibnamefont{Aharonov}},
  \bibinfo{author}{\bibfnamefont{Z.}~\bibnamefont{Landau}}, \bibnamefont{and}
  \bibinfo{author}{\bibfnamefont{J.}~\bibnamefont{Makowsky}},
  \eprint{quant-ph/0611156}.

\bibitem[{\citenamefont{Yoran and Short}()}]{Yoran06}
\bibinfo{author}{\bibfnamefont{N.}~\bibnamefont{Yoran}} \bibnamefont{and}
  \bibinfo{author}{\bibfnamefont{A.~J.} \bibnamefont{Short}},
  \eprint{quant-ph/0706.0872}.

\bibitem[{\citenamefont{Browne}(2007)}]{Browne06}
\bibinfo{author}{\bibfnamefont{D.~E.} \bibnamefont{Browne}},
  \bibinfo{journal}{New J. Phys.} \textbf{\bibinfo{volume}{9}},
  \bibinfo{pages}{146} (\bibinfo{year}{2007}).

\bibitem[{\citenamefont{Lyakhov et~al.}(2007)\citenamefont{Lyakhov, Braun, and
  Bruder}}]{Lyakhov07}
\bibinfo{author}{\bibfnamefont{A.~O.} \bibnamefont{Lyakhov}},
  \bibinfo{author}{\bibfnamefont{D.}~\bibnamefont{Braun}}, \bibnamefont{and}
  \bibinfo{author}{\bibfnamefont{C.}~\bibnamefont{Bruder}},
  \bibinfo{journal}{Physical Review A (Atomic, Molecular, and Optical Physics)}
  \textbf{\bibinfo{volume}{76}}, \bibinfo{pages}{022321}
  (\bibinfo{year}{2007}).

\bibitem[{\citenamefont{Boyer et~al.}(1996)\citenamefont{Boyer, Brassard,
  Hoyer, and Tapp}}]{Boyer96}
\bibinfo{author}{\bibfnamefont{M.}~\bibnamefont{Boyer}},
  \bibinfo{author}{\bibfnamefont{G.}~\bibnamefont{Brassard}},
  \bibinfo{author}{\bibfnamefont{P.}~\bibnamefont{Hoyer}}, \bibnamefont{and}
  \bibinfo{author}{\bibfnamefont{A.}~\bibnamefont{Tapp}}, in
  \emph{\bibinfo{booktitle}{Proceedings of the 4th Workshop on Physics and
  Computation --- PhysComp'96}} (\bibinfo{year}{1996}).

\bibitem[{\citenamefont{Nielsen and Chuang}(2000)}]{Nielsen00}
\bibinfo{author}{\bibfnamefont{M.~A.} \bibnamefont{Nielsen}} \bibnamefont{and}
  \bibinfo{author}{\bibfnamefont{I.~L.} \bibnamefont{Chuang}},
  \emph{\bibinfo{title}{Quantum Computation and Quantum Information}}
  (\bibinfo{publisher}{Cambridge University Press}, \bibinfo{year}{2000}).

\bibitem[{\citenamefont{Arnaud and Braun}(2007)}]{Arnaud07}
\bibinfo{author}{\bibfnamefont{L.}~\bibnamefont{Arnaud}} \bibnamefont{and}
  \bibinfo{author}{\bibfnamefont{D.}~\bibnamefont{Braun}},
  \bibinfo{journal}{Physical Review A (Atomic, Molecular, and Optical Physics)}
  \textbf{\bibinfo{volume}{75}}, \bibinfo{pages}{062314}
  (\bibinfo{year}{2007}).

\end{thebibliography}

\end{document}